\documentclass[twocolumn,numberedappendix,iop]{openjournal}
\usepackage{graphicx,amsmath,amssymb,amstext}
\usepackage{amsbsy,amsfonts,amsthm,color}
\usepackage[colorlinks,linkcolor=blue,citecolor=blue,urlcolor=blue ]{hyperref}
\usepackage[utf8]{inputenc}
\usepackage{float}
\usepackage{xcolor}
\usepackage{ulem}
\usepackage[T1]{fontenc}
\usepackage[title]{appendix}
\usepackage{ulem} 
\usepackage{lineno}

\newcommand{\bq}{\boldsymbol q}
\newcommand{\bx}{\boldsymbol x}

\newcommand{\bk}{\boldsymbol k}
\newcommand{\mo}{\mathcal{O}}

\begin{document}

\title{Validation of the DESI-DR1 $3 \times 2$-pt analysis: scale cut and shear ratio tests \vspace{-4em}}

\author{
N.~Emas,$^{1,*}$
A.~Porredon,$^{2,3,4,5}$
C.~Blake,$^{1}$
J.~DeRose,$^{6}$
J.~Aguilar,$^{7}$
S.~Ahlen,$^{8}$
D.~Bianchi,$^{9,10}$
D.~Brooks,$^{11}$
F.~J.~Castander,$^{12,13}$
T.~Claybaugh,$^{7}$
A.~Cuceu,$^{7}$
A.~de la Macorra,$^{14}$
A.~Dey,$^{15}$
B.~Dey,$^{16,17}$
P.~Doel,$^{11}$
S.~Ferraro,$^{7,18}$
J.~E.~Forero-Romero,$^{19,20}$
C.~Garcia-Quintero,$^{21,22}$
E.~Gaztañaga,$^{12,23,13}$
S.~Gontcho A Gontcho,$^{7,24}$
G.~Gutierrez,$^{25}$
S.~Heydenreich,$^{26}$
K.~Honscheid,$^{27,28,5}$
D.~Huterer,$^{29,30}$
M.~Ishak,$^{31}$
S.~Joudaki,$^{2}$
R.~Joyce,$^{15}$
E.~Jullo,$^{32}$
S.~Juneau,$^{15}$
R.~Kehoe,$^{33}$
D.~Kirkby,$^{34}$
T.~Kisner,$^{7}$
A.~Kremin,$^{7}$
A.~Krolewski,$^{35,36,37}$
O.~Lahav,$^{11}$
M.~Landriau,$^{7}$
J.~U.~Lange,$^{38}$
L.~Le~Guillou,$^{39}$
A.~Leauthaud,$^{26,40}$
M.~Manera,$^{41,42}$
R.~Miquel,$^{43,42}$
S.~Nadathur,$^{23}$
W.~J.~Percival,$^{35,36,37}$
F.~Prada,$^{44}$
G.~Rossi,$^{45}$
R.~Ruggeri,$^{46}$
E.~Sanchez,$^{2}$
C.~Saulder,$^{47}$
A.~Semenaite,$^{1}$
H.~Seo,$^{48}$
J.~Silber,$^{7}$
D.~Sprayberry,$^{15}$
Z.~Sun,$^{49}$
G.~Tarl\'{e},$^{30}$
B.~A.~Weaver,$^{15}$
R.~H.~Wechsler,$^{50,51,52}$
and R.~Zhou$^{7}$ \\
{\it (Affiliations can be found after the references)}
}
\thanks{$^*$E-mail: nemas@swin.edu.au}

\begin{abstract}
Combined survey analyses of galaxy clustering and weak gravitational lensing ($3 \times 2$-pt studies) will allow new and accurate tests of the standard cosmological model.  However, careful validation is necessary to ensure that these cosmological constraints are not biased by uncertainties associated with the modelling of astrophysical or systematic effects.  In this study we validate the combined $3 \times 2$-pt analysis of the Dark Energy Spectroscopic Instrument Data Release 1 (DESI-DR1) spectroscopic galaxy clustering and overlapping weak lensing datasets from the Kilo-Degree Survey (KiDS), the Dark Energy Survey (DES), and the Hyper-Suprime-Cam Survey (HSC).  By propagating the modelling uncertainties associated with the non-linear matter power spectrum, non-linear galaxy bias and baryon feedback, we design scale cuts to ensure that measurements of the matter density and the amplitude of the matter power spectrum are biased  by less than 30\% of the statistical error.  We also test the internal consistency of the data and weak lensing systematics by performing new measurements of the lensing shear ratio.  We demonstrate that the DESI-DR1 shear ratios can be successfully fit by the same model used to describe cosmic shear correlations, and analyse the additional information that can be extracted about the source redshift distributions and intrinsic alignment parameters.  This study serves as crucial preparation for the upcoming cosmological parameter analysis of these datasets.
  \\[1em]
  \textit{Keywords:} Weak gravitational lensing, large-scale structure of Universe, Cosmological parameters
\end{abstract}

\maketitle

\section{Introduction}

We have entered the era of precision cosmology, in which the parameters that define the cosmological model, such as its geometry, composition and growth of structure over time, may be accurately measured by cosmological probes.  Combinations of multiple observational probes are particularly useful in this regard, to test and improve our knowledge of cosmological models, mitigate systematic errors, learn the astrophysical properties of the observables, and break the degeneracies between parameters.

In particular, the combination of weak gravitational lensing, the correlated distortions of galaxy shapes as their light streams through the cosmic web, and galaxy redshift surveys, mapping out the large-scale structure of the Universe, allow us to measure the amplitude of matter clustering through parameters such as $S_8 \equiv \sigma_8 \sqrt{\Omega_\mathrm{m}/0.3}$, where $\sigma_8$ is the amplitude of matter fluctuations on a scale of 8 $h^{-1}$Mpc, and $\Omega_\mathrm{m}$ is the matter density parameter.  The current generation of cosmological facilities are expected to provide stringent tests of the cosmological model, including accurate values of parameters such as $S_8$.  However, the advent of these new datasets should be accompanied by rigorous attention to modelling, data analysis, error estimation, and likelihood determination \citep[e.g.][]{2021A&A...646A.129J, 2021arXiv210513548K}.

In this study we focus on the combination of spectroscopic data from the Dark Energy Spectroscopic Instrument Data Release 1 \citep[DESI-DR1,][]{DESI1, DESI2, DESI3, DESI4, 2025JCAP...07..028A}, and current weak lensing datasets.  DESI is the largest current spectroscopic galaxy survey, and aims to understand the nature of dark energy by measuring the distribution of at least 40 million galaxies across $14{,}000$ deg$^2$ during an initial nominal five-year survey period.  DESI is attached to the 4-metre Mayall telescope at Kitt Peak Observatory, Arizona, and uses 5000 optical fibres to achieve a high multiplex of galaxy observations.  DESI-DR1 is the data assembly corresponding to roughly the first full year of survey operations \citep{2025arXiv250314745D}.

In \cite{2025arXiv251215960P}, we present a cosmological analysis combining DESI-DR1 data with shear catalogues from three current weak lensing surveys: the Kilo-Degree Survey \citep[KiDS-1000,][]{2021A&A...645A.105G}, the Dark Energy Survey Year 3 \citep*[DES-Y3,][]{2021MNRAS.504.4312G}, and the Hyper-Suprime-Cam Survey Year 1 \citep[HSC-Y1,][]{2018PASJ...70S..25M} and Year 3 \citep[HSC-Y3,][]{HSC-Y31} datasets.  DESI overlaps with all these weak lensing surveys, allowing us to perform auto- and cross-correlation measurements between foreground galaxy positions and the shapes of background sources.  These correlations include cosmic shear (source-source), galaxy clustering (lens-lens) and galaxy-galaxy lensing (source-lens), which we collectively refer to as a $3 \times 2$-pt analysis \citep{2018MNRAS.474.4894J, 2018PhRvD..98d3526A, 2021A&A...646A.140H, 2022PhRvD.105b3520A, 2023PhRvD.108l3521S, 2023arXiv230400704M}.  In our study, which is a companion to \cite{2025arXiv251215960P}, we focus on the modelling validation of this $3 \times 2$-pt analysis: that is, the tests we undertake to ensure that systematic errors associated with the choice of cosmological or astrophysical modelling, or other analysis choices such as scale or source-lens cuts, are sub-dominant to the statistical error budget determined by sample variance and measurement noise.

Generally speaking, three validation approaches are available for $3 \times 2$-pt analyses.  The first approach is to use large-scale cosmological simulations to test cosmological parameter recovery, modelling accuracy and data covariance.  The results of our large-scale ``mock challenge'' have already been presented by \cite{2024arXiv241212548B}; results are subject to the limitations of how accurately a simulation can represent all effects present in the data.  The second technique, which we focus on in this study, is to assess ``analytically'' the sensitivity of cosmological parameters to model assumptions such as non-linear matter power spectrum, non-linear galaxy bias, and baryon feedback modelling \citep[e.g.,][]{2017arXiv170609359K, 2021arXiv210513548K, 2022PhRvD.105b3514A, DES_cut, 2023OJAp....6E..36D}.  By propagating reasonable variations in these modelling assumptions to variations in cosmological parameters, we design scale cuts to minimize the effect of these analysis choices, rendering our final measurements more robust.  The third validation approach is to consider variations in the fitting process of real data; however, since the data is ``blinded'' to prevent confirmation bias, this approach is not available for analysis design in advance.

A complementary method for validating weak lensing analyses, which we also adopt in this study, is to use the shear ratio (SR) test \citep{Jain&Taylor, Bernstein&Jain2004, Taylor2007, Kitching2008, 2016A&A...592L...6S, 2021A&A...645A.105G, 2022PhRvD.105h3529S, Emas_SR}. This method compares the tangential shear signal originating from different source tomographic samples around the same lens sample.  The shear ratio can be used to validate the internal consistency of the data and help identify systematic biases associated with the source redshift distribution, intrinsic alignment parameters, and multiplicative shear bias parameters.  This part of our analysis extends previous work \citep{Emas_SR} using data from the Baryon Oscillation Spectroscopic Survey (BOSS) as the lenses for a shear ratio test.  In \cite{Emas_SR} we found that combining SR with 2-pt cosmic shear statistics can be an alternative to using informative priors in the mean redshifts of the source galaxies, and helpful for improving the measurements of cosmological and weak lensing astrophysical parameters.  In our current study we also apply a shear ratio test using the combination of DESI and weak lensing surveys, to check the sufficiency of our cosmological and astrophysical modelling.

Our paper is structured as follows.  In Section \ref{sec:data} we introduce the datasets we analyse, namely the DESI-DR1 spectroscopic catalogue and the weak lensing data from KiDS-1000, DES-Y3, HSC-Y1, and HSC-Y3.   We describe our measurements, covariance, modelling, and parameterization in Section \ref{sec:methods}.   We present the scale cut validation in Section \ref{sec:Model_valid}, and the shear ratio test in Section \ref{sec:SR_test}.  Finally, we summarize our conclusions in Section \ref{sec:conc}.

\section{Data}
\label{sec:data}

\begin{figure*}
  \includegraphics[width=\linewidth]{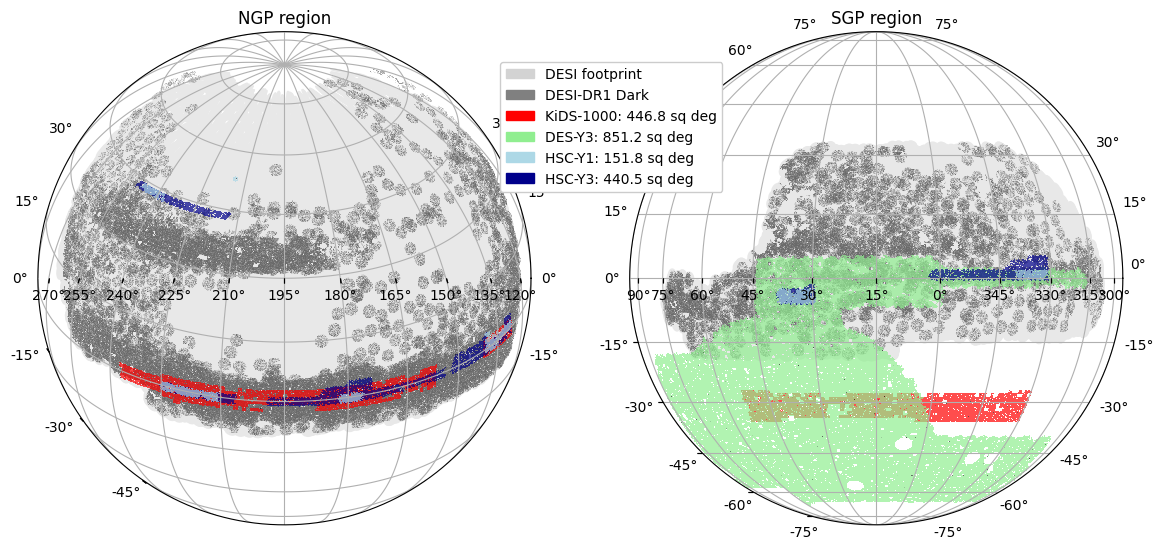}
  \caption{The footprints on the sky of weak lensing surveys: KiDS-1000 (red), DES-Y3 (green), HSC-Y1 (light blue), and HSC-Y3 (dark blue), and their overlap with the DESI-DR1 dataset (dark grey). The two hemispheres are centred on the North Galactic Pole (NGP) and South Galactic Pole (SGP) regions.  The overall footprint of the DESI survey is shaded as light grey.}
  \label{fig:map}
\end{figure*}

In this section we present a brief overview of the galaxy catalogues used in our analysis: the DESI-DR1 spectroscopic clustering catalogue, and the KiDS, DES and HSC weak lensing imaging shape catalogues.  The sky distributions of DESI and the weak lensing surveys, illustrating their overlaps, is displayed in Figure~\ref{fig:map}.  These datasets are briefly summarised below.

\subsection{DESI-DR1}

The Dark Energy Spectroscopic Instrument (DESI) is a stage-IV ground-based dark energy experiment that studies cosmic expansion through baryon acoustic oscillations (BAO) and the growth of structure through redshift-space distortions (RSD) using wide-area galaxy and quasar redshift surveys \citep{DESI1, DESI2, DESI3, DESI4, 2025JCAP...07..028A}. DESI is installed at the prime focus of the 4-m Mayall telescope in Kitt Peak, Arizona.  The instrument contains an optical corrector \citep{2024AJ....168...95M} and robotic fibres \citep{2024AJ....168..245P} capable of observing up to $5{,}000$ spectra simultaneously across a wavelength range of 360 to 980 nm, with a corresponding spectroscopic resolution increasing from $\sim 2000$ to $\sim 5000$ \citep{2023AJ....165....9S}, with the data processed by the DESI spectroscopic pipeline \citep{2023AJ....165..144G}.  In its nominal phase, DESI will observe over 40 million galaxy redshifts, split into different components including a Bright Galaxy Survey (BGS), Luminous Red Galaxies (LRG), Emission Line Galaxies (ELG), and Quasi-Stellar Objects (QSO), which are selected from the DESI Legacy optical imaging surveys \citep{2019AJ....157..168D}.  The overall DESI observing strategy is summarised by \cite{2023AJ....166..259S}.

\begin{figure}
  \includegraphics[width=\linewidth]{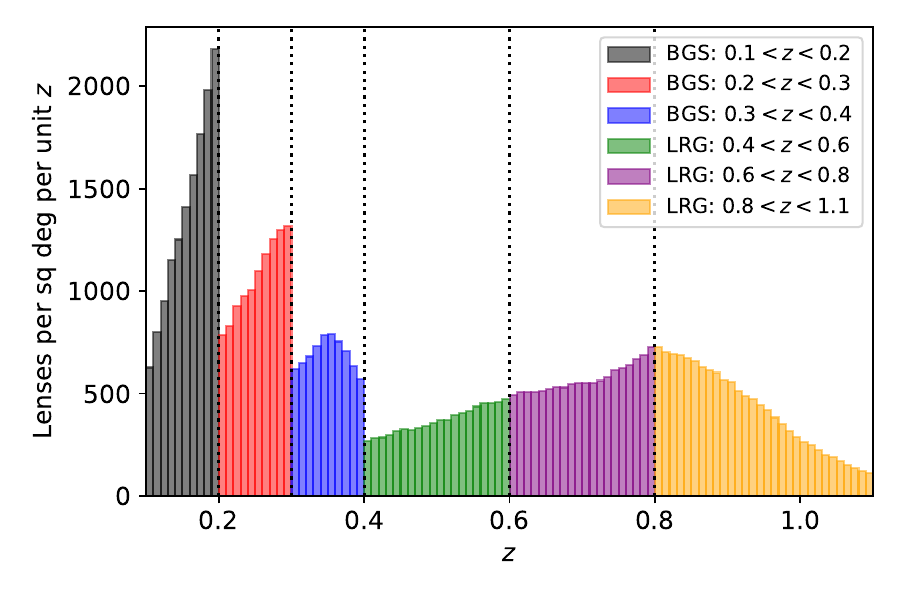}
  \caption{The average angular density per unit redshift of the DESI-DR1 lens samples used in our analysis, spanning three Bright Galaxy Survey (BGS) and three Luminous Red Galaxy (LRG) tomographic bins.}
  \label{fig:nz_lens}
\end{figure}

In our lensing analysis we use the subset of BGS \citep{BGS1, BGS2} and LRG \citep{LRG1,LRG2} galaxies that are included in the DESI Data Release 1 \citep{2025arXiv250314745D}.   We divide these lenses into six tomographic bins.  The three first bins use BGS tracers with redshift ranges $[0.1, 0.2, 0.3, 0.4]$, and the three last bins use LRG tracers with redshift divisions $[0.4, 0.6, 0.8, 1.1]$.  The angular density per unit redshift of the lens samples, as a function of redshift, is displayed in Figure~\ref{fig:nz_lens}.  These lenses overlap with the three weak lensing surveys we use in our analysis: the KiDS, DES and HSC lensing surveys.  We note that for our BGS galaxy-galaxy lensing measurements, we use a fainter luminosity cut in each redshift range than applied in the DESI ``Key Project'' analysis in order to reduce the noise in these measurements, as discussed by \cite{2024MNRAS.533..589Y} and \cite{SvenDESI}.

\subsection{KiDS-1000}

\begin{figure*}
  \includegraphics[width=\linewidth]{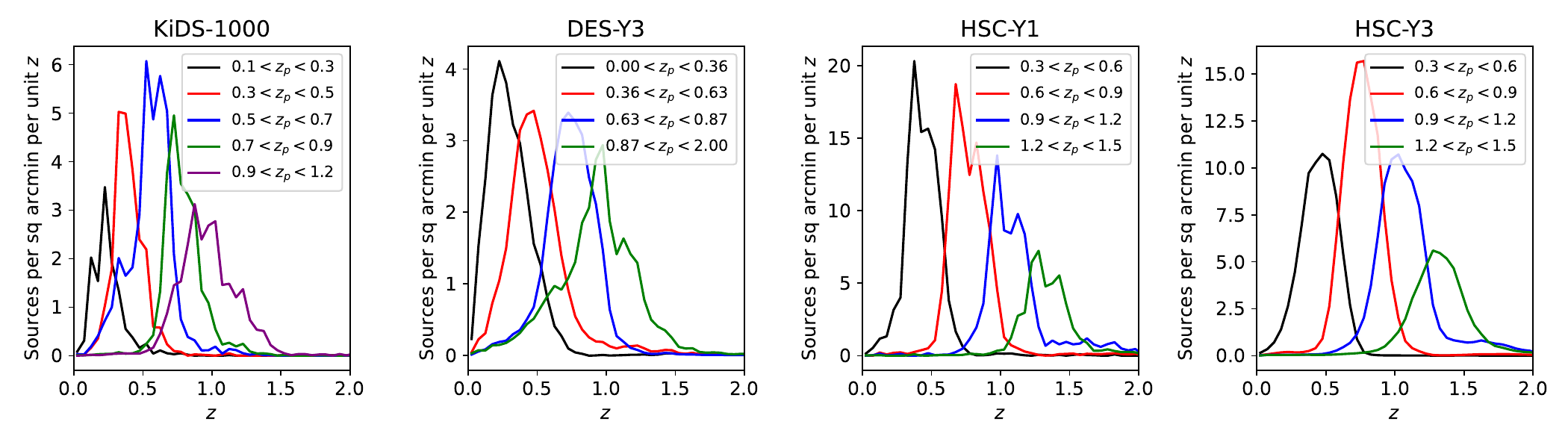}
  \caption{The weighted angular density per unit redshift of the weak lensing source datasets used in our analysis, spanning different tomographic samples for the KiDS-1000, DES-Y3, HSC-Y1 and HSC-Y3 datasets.  The ``spikes'' in the redshift distributions, which appear for some surveys, are caused by sample variance when these distributions are estimated from calibration samples which only span small areas of sky.}
  \label{fig:nz_source}
\end{figure*}

The Kilo-Degree Survey is an optical wide-field imaging project designed to study weak gravitational lensing.  KiDS uses the OmegaCAM instrument on the Very Large Telescope (VLT) Survey Telescope at the European Southern Observatory's Paranal Observatory, and images the sky in optical filters $u, g, r, i$.  KiDS targets two main sky regions, KiDS-North and KiDS-South, with a total planned coverage of $1{,}350$ deg$^2$.  Complementary imaging in the near-infrared bands $Z, Y, J, H, K_s$ is provided by the VISTA-VIKING survey, resulting in deep, wide, nine-band imaging datasets with improved photometric redshift calibration \citep{2021A&A...647A.124H}.  In this project we use the KiDS-1000 weak lensing catalogue \citep{2021A&A...645A.105G}, which is part of the fourth public data release \citep{2019A&A...625A...2K} and includes 1006 deg$^2$ of lensing data.  The shape of the sources in this catalogue is measured using the {\it lens}fit model fitting technique \citep{miller_lensfit, miller_lensfit2,conti_lensfit}. The area of overlap of KiDS-1000 and DESI-DR1 is 446.8 deg$^2$.  Our analysis uses the five tomographic source bins defined by KiDS, which are divided by photometric redshifts $z_B = [0.1, 0.3, 0.5, 0.7, 0.9, 1.2]$, where the redshift distribution of the sources is calibrated as described by \cite{2021A&A...647A.124H}.  The source redshift distributions for each of the tomographic samples are illustrated for KiDS and the other weak lensing surveys in Figure~\ref{fig:nz_source}.  Cosmological fits to the KiDS-1000 dataset are presented by \cite{2021A&A...646A.140H}.  We note that recent results from the latest data assembly, KiDS-Legacy \citep{2025A&A...703A.158W}, incorporate improvements in redshift distribution estimation and calibration, and show consistent $S_8$ constraints with those from Planck.

\subsection{DES-Y3}

The Dark Energy Survey has conducted the largest existing weak lensing survey.  DES utilizes the Dark Energy Camera mounted on the Blanco 4-metre Telescope and uses five broadband filters ($g, r, i, z, Y$) \citep{DES_2}. For this analysis, we use data from the first three years of DES operation (Y3), which were collected between 2013 and 2016 over a $4{,}143$ deg$^2$ area of sky \citep*{DES_obs_paper}.  The DES shear catalogue \citep{2021MNRAS.504.4312G} is created using the self-calibrating shear measurement pipeline \textsc{metacalibration} \citep{Metacalibration1, Metacalibration2} and contains over $10^8$ objects.  The DES-Y3 catalogue has an overlap of 851.3 deg$^2$ with the DESI-DR1 footprint.  We analysed the DES-Y3 sources in the four tomographic bins used by the collaboration, which are divided by photometric redshifts $[0.00, 0.36, 0.63, 0.89, 2.00]$, as detailed in \citet*{DES_nz}.  The DES collaboration has performed validations and cosmological fits based on these catalogues \citep*{2022PhRvD.105b3520A, 2022PhRvD.105b3514A, DES_cut}. We note that we use the new version of the DES-Y3 catalogue with the updated source tomographic bin assignments, as described in footnote 5 of \cite{DESblueshear}.

\subsection{HSC-Y1 and HSC-Y3}

The Subaru Hyper-Suprime-Cam survey is the deepest large-scale weak lensing survey currently available.  HSC uses the 8.2-metre Subaru Telescope to map a planned 1400 deg$^2$ of the sky divided into six regions with $g, r, i, z, Y$ filters, where the $i$-band seeing is around $0.6\arcsec$.  For this analysis, we utilise the first-year data release \citep[Y1,][]{2018PASJ...70S..25M} and the third-year data release \citep[Y3,][]{HSC-Y31} of the HSC weak lensing survey.  The main difference between the catalogues lies in the area covered, where the overlap area with DESI-DR1 is 151.8 deg$^2$ and 440.5 deg$^2$ for HSC-Y1 and HSC-Y3, respectively.  The unweighted source densities of the two data releases are $(24.6, 22.9)$ arcmin$^{-2}$.  For calibration, HSC-Y1 uses a re-Gaussianization PSF correction method \citep{2018PASJ...70S..25M}, while HSC-Y3 incorporates realistic image simulations that emulate survey observation conditions and are based on galaxy images from the Hubble Space Telescope (HST) COSMOS field \citep{HSC-Y31}.  Our work uses the four tomographic bins of sources defined by the HSC collaboration, with a photometric redshift divisions $[0.3, 0.6, 0.9, 1.2, 1.5]$.  Cosmological fits to these datasets are presented by \cite{2019PASJ...71...43H, hsc_cut, HSCY3-cosmic, 2023PhRvD.108l3519D}.

\section{Combined-probe methodology}
\label{sec:methods}

\subsection{$3 \times 2$-pt correlation functions}

We use the lens and source galaxy catalogues to analyse the statistics of two cosmological fields which trace the underlying matter distribution: the galaxy density field and the weak lensing shear field.  We capture the information in these two fields by measuring 2-point correlations as a function of separation.  The auto- and cross-correlations between the fields provide three types of correlation functions, which are known as the $3 \times 2$-pt correlations.

The first type of correlation function we include in our study is cosmic shear.  Cosmic shear is the correlated distortion between galaxy shapes imprinted as source galaxy light passes through the matter distribution in the cosmic web.  The 2-point correlation function between the galaxy shears from two source tomographic redshift bins $\{ i,j \}$ is represented by the function $\xi^{i,j}_\pm(\theta)$ as a function of angular separation $\theta$, where $\pm$ represents the sum and difference of the products of the tangential and cross-components of the projected shear.

In our analysis we utilise the same angular separation bins adopted by each weak lensing survey collaboration for their analysis.  For KiDS-1000 we use 9 logarithmic separation bins in the range $0.5 - 300'$ \citep{2020A&A...633A..69H, kids1000_cut2}.  For DES-Y3 we use 20 separation bins in the range $2.5 - 250'$ \citep{2022PhRvD.105b3514A, DES_cut}.  For HSC-Y1 we use 22 scale bins in the range $10^{0.05} - 10^{2.25}$$'$, where we analyse the same subset of separations $7.1 - 56.2'$ for $\xi_+$ and $28.2 - 178.0'$ for $\xi_-$ as the HSC-Y1 collaboration, motivated by the signal-to-noise ratio and impact of point spread function systematics \citep{hsc_cut}.  For HSC-Y3 we utilise 16 separation bins in the range $2.188-247.749'$ \citep{HSCY3-cosmic}.

Galaxy clustering is the 2-point correlation between the positions of lens galaxies in a given redshift bin, induced by the gravitational growth of cosmic structure.  In our analysis we describe lens galaxy clustering using the projected correlation function $w_{\mathrm{p}}(r_p)$ as a function of transverse comoving separation $r_p$.  Given that DESI has measured spectroscopic redshifts for the lens galaxies in each sample, we can de-project their angular positions at a given redshift.  Hence we use the projected correlation function in preference to the angular galaxy correlation function $w(\theta)$, which would yield a measurement with a lower signal-to-noise ratio in a given redshift bin.  However, the projected correlation function still suppresses the information along the line-of-sight, which would be affected by redshift-space distortions (i.e., galaxy peculiar velocities), the modelling of which is beyond the scope of our current analysis.

Our measurement of the projected correlation functions of the DESI lens samples is presented by \cite{SvenDESI}, and we briefly summarise the key details here.  The projected correlation function is determined from the 3D galaxy correlation function $\xi_{gg}$ as,
\begin{equation}
    w_{\mathrm{p}}(r_{\mathrm{p}}) = 2 \int^{r_{\pi,\mathrm{max}}}_{0} \xi_{\rm gg}(r_\mathrm{p},r_{\pi}) \, dr_{\pi},
\label{eq_galclus}
\end{equation}
where $r_p$ and $r_\pi$ are the transverse and line-of-sight separations between galaxy pairs, and $r_{\pi,\mathrm{max}}$ is the upper bound of the integration along the line-of-sight direction.  We apply Equation~\ref{eq_galclus} as a sum over line-of-sight separation bins, and estimate $\xi_{gg}$ using the Landy-Szalay estimator \citep{1993ApJ...412...64L}:
\begin{equation}
    \xi_{\rm gg}(r_\mathrm{p},r_{\pi}) = \frac{\rm DD - 2 DR + RR}{\rm RR},
\end{equation}
where $DD$, $RR$, and $DR$ are the normalized number of galaxy-galaxy, random-random, and galaxy-random pairs within bins of projected separation $r_p$ and line-of-sight separation $r_{\pi}$.

We perform the projected correlation function measurement in 15 logarithmic separation bins in the range $0.08 < r_p < 80 \, h^{-1}$Mpc, where we will later exclude small separations which present modelling difficulties, as discussed below.  We adopt $r_{\pi,\mathrm{max}} = 100 \, h^{-1}$Mpc for our analysis. This choice is larger than redshift-space distortion (RSD) offsets, such that Finger of God (FoG) effects do not significantly affect the $w_{\rm p}$ measurement.  As we discuss below, we also apply a minimum projected separation cut in the clustering analysis, which serves to remove non-linear RSD effects.  As discussed by \cite{SvenDESI}, our correlation function measurements include weights which compensate for target selection systematics, spectroscopic incompleteness and fibre collisions \citep{2025JCAP...07..017A}.

The final 2-point correlation we consider is galaxy-galaxy lensing, which we quantify as the average tangential shear of background galaxies in source samples $j$ around foreground galaxies in lens samples $i$ as a function of angular separation, denoted by $\gamma^{i,j}_t(\theta)$.  Our galaxy-galaxy lensing measurements using DESI lens galaxies and KiDS, DES and HSC source samples are described by \cite{SvenDESI}, and briefly summarised below.  We use an estimator that combines the catalogues of source galaxies ($\mathrm{s}$), lens galaxies ($\mathrm{l}$), and an unclustered (random) lens catalogues ($\mathrm{r}$) that shares the same selection function as the lens galaxies:
\begin{equation}\label{gammat}
    \hat{\gamma}_\mathrm{t}(\theta) = \frac{\sum\limits_{\mathrm{ls}} w_\mathrm{l} w_\mathrm{s} \, e_\mathrm{{t,ls}}}{\sum\limits_\mathrm{{ls}} w_\mathrm{l} w_\mathrm{s}} - \frac{\sum\limits_\mathrm{{rs}} w_\mathrm{r} w_\mathrm{s} \, e_\mathrm{{t,rs}} }{\sum\limits_\mathrm{{rs}} w_\mathrm{r} w_\mathrm{s}} ,
\end{equation}
where $w$ represents the weight of each object, and $e_\mathrm{t}$ is the tangential ellipticity of the source, projected onto an axis perpendicular to the line connecting the source and data or random lens.  Hence, we do not apply the ``boost'' correction which aims to quantify physical source-lens associations not caused by gravitational lensing, which can occur in cases where the source and lens redshift distributions partially overlap, although we note that this correction is negligible for the range of scales included in our analysis \citep{2024OJAp....7E..57L}.   We measured the average tangential shear in 15 logarithmically-spaced separation bins in the range $0.3 - 300'$.  Later, we will exclude measurements at small separations which may be difficult to model.

\subsection{Covariance}
\label{Covariance}

The covariance of the data vector describes the correlated uncertainties in the correlation function measurements across separation bins, lens and source tomographic samples, and type of correlation function.  Denoting the data vector by $\mathbf{D}$, which represents a concatenation of the $3 \times 2$-pt correlation functions, the covariance between two elements $i$ and $j$ is defined by,
\begin{equation}
    C_{ij} = \langle D_i \, D_j \rangle - \langle D_i \rangle \, \langle D_j \rangle ,
\label{eq:cov}
\end{equation}
where the angled brackets $\langle ... \rangle$ denote an average over a statistical ensemble of realisations.

In our study we use a Gaussian analytical approach to determine the data covariance, which includes sample variance, noise and mixed terms, as well as super-sample variance.  Our implementation of these analytical terms is presented by \cite{2024arXiv241212548B}, where our computations follow the methods of \cite{2017MNRAS.470.2100K} and \cite{2021A&A...646A.129J}, with modifications to treat the case of $w_{\mathrm{p}}(r_{\mathrm{p}})$.  \cite{2024arXiv241212548B} includes comparisons between the analytical covariance and a numerical covariance which evaluates Equation~\ref{eq:cov} using a representative set of mock catalogues, demonstrating that the analytical covariance is accurate within an uncertainty of $\sim 10\%$.  We do not include the non-Gaussian contribution to the covariance, which
is currently negligible for the scales and cases of interest \citep{2021A&A...646A.129J}.  The galaxy-galaxy lensing section of the covariance is corrected for boundary effects using an ensemble of fast mocks matched to the footprint of the lensing surveys, as discussed and tested by \cite{2024MNRAS.533..589Y}.

\subsection{Correlation models}
\label{sec:models}

We follow the approach of the DES-Y3 analysis \citep*{2022PhRvD.105b3520A, 2022PhRvD.105h3529S} for computing the theoretical models of the $3 \times 2$-pt correlations, which we summarise in this section.

\subsubsection{Cosmic shear}

For the cosmic shear correlation function $\xi^{i,j}_\pm(\theta)$ between two redshift bins $i$ and $j$, the angular 2-point correlation function is expressed in terms of the convergence power spectrum  $C_{\kappa}(\ell)$ at an angular wavenumber $\ell$ as,
\begin{equation}
    \xi^{i,j}_{\pm}(\theta) = \sum_{\ell} \frac{2\ell + 1}{4\pi} G_{\ell}^{\pm}(\cos\theta) 
\left[ C_{\kappa, \text{EE}}^{i,j}(\ell) \pm C_{\kappa, \text{BB}}^{i,j}(\ell) \right] ,
\label{eq:xi_model}
\end{equation}
where $G_{\ell}^{\pm}(x)$ are computed from the Legendre polynominals $P_{\ell}(x)$ and averaged over the angular bin \citep[see][]{2017arXiv170609359K, 2022PhRvD.105b3520A}. For cosmological lensing, $C_{\kappa, \text{BB}} = 0$ and the B-mode contribution originates from intrinsic alignments.  We describe the model we adopt for intrinsic alignments in Section \ref{sec:iamodel}.

The angular cross-power spectra $C_{\kappa}^{i,j}(\ell)$ can be derived from the non-linear 3D matter power spectrum $P_{mm}$ using the flat-sky and Limber approximations \citep{Limber,Limber2} as, 
\begin{equation}
C_{\kappa}^{i,j}(\ell) = \int_0^{\chi_H} d\chi \, \frac{W^i_\kappa(\chi) W^j_\kappa(\chi)}{\chi^2} 
P_{mm} \left( k = \frac{\ell + 0.5}{\chi} , z(\chi) \right) ,
\label{eq:Limber2}
\end{equation}
where $\chi$ is the comoving angular diameter distance, $\chi_H$ is the horizon distance and the lensing efficiency kernel $W_i(\chi)$ for the tomographic bin $i$ is defined by,
\begin{equation}
W^{i}_\kappa (\chi) = \frac{3H_0^2 \Omega_\mathrm{m}}{2c^2} \frac{\chi}{a(\chi)} 
\int_{\chi}^{\chi_H} d\chi' \, n^{i}_\mathrm{s}(\chi') \frac{\chi' - \chi}{\chi'} ,
\label{eq:kernel_v2}
\end{equation}
where $n^i_\mathrm{s}(\chi)d\chi$ is the effective number of source galaxies in the interval $d\chi$ in the $i$-th tomographic bin, $a(\chi)$ is the cosmic scale factor, $H_0$ is the Hubble constant, $\Omega_{\rm m}$ is the density of matter, and $c$ is the speed of light. We calculate the matter power spectrum, $P_{mm}$ with \textsc{camb} \citep{2000ApJ...538..473L} and we use \textsc{hmcode2020} with baryon feedback \citep{hmcode2020} as our baseline model for the non-linear matter power spectrum.  For the shear ratio test, we also consider two alternative non-linear models: \textsc{hmcode2020} without baryon feedback and \textsc{halofit} \citep{Takahashi_2012}.

\subsubsection{Galaxy-galaxy lensing}

For galaxy-galaxy lensing, the model for the average tangential shear of source galaxies in tomographic bin $j$ around lens galaxies in redshift bin $i$ is,
\begin{equation}
\gamma^{i,j}_\mathrm{t}(\theta) = \sum \limits_{\ell} \frac{2\ell+1}{4\pi \ell(\ell+1)} {P^2_{\ell}}(\cos\theta) C^{i,j}_{g\kappa,\mathrm{tot}}(\ell) ,
\label{eq:gammat_model_v2}
\end{equation}
where  $\ell$ is the 2D multipole moment, $C^{i,j}_{g\kappa,\mathrm{tot}}(\ell)$ is the total angular galaxy-matter cross-power spectrum \citep*{2022PhRvD.105b3520A, 2022PhRvD.105h3529S}, and ${P^2_{\ell}}(\cos\theta)$ is computed from the Legendre polynomials and averaged over the angular bin \citep[see][]{2017arXiv170609359K}.  The total cross-spectrum includes effects originating from galaxy density, intrinsic alignment and magnification, and can be written as,
\begin{equation}
\label{eq_14}
C^{i,j}_{g\kappa,\mathrm{tot}}=C^{i,j}_{g\kappa}+C^{i,j}_{g\kappa,\mathrm{IA}}+C^{i,j}_{g\kappa,\mathrm{mag}}+C^{i,j}_{g\kappa,\mathrm{IA} \times \mathrm{mag}},
\end{equation}
where $C^{i,j}_{g\kappa}$ is the cross-power spectrum associated with cosmic shear, $C^{i,j}_{g\kappa,\mathrm{IA}}$ is the contribution due to intrinsic alignment, $C^{i,j}_{g\kappa,\mathrm{mag}}$ is due to lens magnification, and $C^{i,j}_{g\kappa,\mathrm{IA} \times \mathrm{mag}}$ is the cross-term between lens magnification and intrinsic alignment (see \cite*{2022PhRvD.105h3529S} for the full relation).

The average tangential shear $\gamma_\mathrm{t}$ is a non-local measurement of the galaxy-mass correlation, which can receive contributions from the highly non-linear distribution of galaxy mass, even at relatively large angular scales. To address this issue, we include analytic marginalization over the mass enclosed within these angular scales, following the approach proposed by \cite{2020MNRAS.491.5498M}. They introduce additional nuisance parameters associated with a point-mass term, which modifies the inverse covariance to facilitate the analytic marginalization.  When performing our scale-cut validation, we use the point-mass marginalization for $\gamma_\mathrm{t}$.

\subsubsection{Galaxy clustering}

In the galaxy clustering model, the correlation function multipoles $\xi_{\ell}(s)$ are obtained as a Hankel transform of the power spectrum multipoles as,
\begin{equation}
    \xi_\ell(s) = \frac{1}{2\pi^2} \int k^2 \, P_\ell(k) \, j_\ell(ks) \, dk ,
\label{eq:hankel}
\end{equation}
where $s = \sqrt{r_p^2 + r_\pi^2}$ is the separation between two galaxies, $j_\ell(ks)$ is the spherical Bessel function of order $\ell$, and $P_\ell(k)$ are the power spectrum multipoles defined by,
\begin{equation}
    P_\ell(k) = \frac{2\ell +1}{2} \int^1_{-1} P_{gg}(k,\mu) \, L_{\ell}(\mu) \, d\mu ,
\end{equation}
where $\mu = \cos{\theta}$ and $L_{\ell}(\mu)$ is a Legendre polynomial. $P_{gg}(k,\mu)$ is the galaxy power spectrum in redshift space using a linear Kaiser model \citep{1987MNRAS.227....1K},
\begin{equation}
\label{Pgg}
    P_{gg} (k, \mu, z_l) = P_{mm}(k,z_l) \left[ b(z_l,k) + f(z_l) \, \mu^2 \right]^2,
\end{equation}
where $b(z_l,k)$ is the linear galaxy bias and $f(z_l)$ is the growth rate of structure.  We use the \textsc{hankl} package \citep{karamanis2021hankl} to evaluate Equation~\ref{eq:hankel} in a fast way, using the FFTlog algorithm. Finally, we substitute our model for $\xi_\ell(s)$ in Equation \ref{eq_galclus} to calculate the projected correlation function, $w_p(r_p)$. 

\begin{table}
\centering
\begin{tabular}{c c c } 
 \hline
 Parameter & Range \\
 \hline
 \textbf{Cosmological parameters}\\ 
 $\Omega_\mathrm{m}$ & [0.1, 0,9]\\
 $h$ & [0.55, 0.91]\\
 $\Omega_\mathrm{b}$ & [0.03, 0.07]\\
 $n_s$ & [0.87, 1.07]\\
 $A_s$ & [$0.5 \times 10^{-9}$, $5.0 \times 10^{-9}$]\\
 \textbf{Systematic parameters}\\ 
 Source redshift: $\Delta z$ & $[-0.2, 0.2]$ \\
 Intrinsic alignment: $A_1$, $\eta_1$  & $[-5.0, 5.0]$ \\
 Shear calibration: $m$ & $[-0.1, 0.1]$ \\  
 Baryon feedback: $\log T_{\rm AGN}$ & [7.3, 8.3]\\
 \hline
 Parameter & Prior \\
 \hline
 \textbf{KiDS-1000} \\
 uncorr $\Delta z_1$ & $\mathcal{N}(0.000, 1.0)$ \\
 uncorr $\Delta z_2$ & $\mathcal{N}(-0.181, 1.0)$ \\
 uncorr $\Delta z_3$ & $\mathcal{N}(-1.110, 1.0)$ \\
 uncorr $\Delta z_4$ & $\mathcal{N}(-1.395, 1.0)$ \\
 uncorr $\Delta z_5$ & $\mathcal{N}(1.265, 1.0)$ \\
 $m_1$ & $\mathcal{N}(-0.009, 0.019)$ \\ 
 $m_2$ & $\mathcal{N}(-0.011, 0.020)$ \\
 $m_3$ & $\mathcal{N}(-0.015, 0.017)$ \\
 $m_4$ & $\mathcal{N}(0.002, 0.012)$ \\
 $m_5$ & $\mathcal{N}(0.007, 0.010)$ \\
 \textbf{DES-Y3} \\
 $\Delta z_i$ & $\mathcal{N}(0,[0.018,0.015,0.011,0.017])$\\
 $m_1$ & $\mathcal{N}(-0.0063, 0.0091)$ \\ 
 $m_2$ & $\mathcal{N}(-0.0198, 0.0078)$ \\
 $m_3$ & $\mathcal{N}(-0.0241, 0.0076)$ \\
 $m_4$ & $\mathcal{N}(-0.0369, 0.0076)$ \\
 \textbf{HSC-Y1} \\
 $\Delta z_i$ & $\mathcal{N}(0 ,[0.0374, 0.0124, 0.0326, 0.0343])$\\
 $m_i$ & $\mathcal{N}(0.0, 0.01)$\\
  \textbf{HSC-Y3} \\
 $\Delta z_i$ ($i=1,2$) & $\mathcal{N}(0 ,[0.024, 0.022])$\\
 $m_i$ & $\mathcal{N}(0.0, 0.01)$\\
 \hline
\end{tabular}
\caption{The set of cosmological, astrophysical and systematic parameters we consider in our analysis.  The top section describes settings which apply to all surveys, and the bottom section of the table lists individual survey specifications. We note that for HSC-Y3 only the first two source bins have Gaussian priors for $\Delta z$, with the last two bins having broad flat priors.}
\label{table:priors}
\end{table}

\subsection{Lensing nuisance parameter model}
\label{sec:iamodel}

We now describe the astrophysical and systematic modelling we use in our analysis.  The parameters and priors we adopt are listed in Table \ref{table:priors}.

\textbf{Galaxy bias}: We assume a linear relation between the lens galaxy density field and the underlying matter density field ($\delta_g = b \, \delta_m $), with bias parameters $b_i$ for lens bin $i$. For galaxy-galaxy lensing, galaxy bias contributes to $C^{i,j}_{g\kappa}$ and $C^{i,j}_{g\kappa, \mathrm{IA}}$ and for galaxy clustering, galaxy bias contributes to the galaxy power spectrum as shown in Equation \ref{Pgg}.  Linear galaxy bias cancels out when modelling the shear ratio.

\textbf{Source redshift distribution}: There are uncertainties in the source redshift probability distribution, reflecting errors in the photometric-redshift calibration method.  We model these uncertainties using offset parameters denoted by $\Delta z_i$ (for source bin $i$) which modify the source redshift distribution such that,
\begin{equation}
    n^{i}_\mathrm{s}(z) = n^{i}_\mathrm{s}(z - \Delta z_{i})
\end{equation}
which alters the lensing efficiency kernel shown in Equation \ref{eq:kernel_v2}.  We perform our analysis both including and excluding the informative Gaussian prior in $\Delta z$ established by each weak lensing collaboration.  When using the Gaussian $\Delta z$ prior, we adopt the DES-Y3 priors from \citet{2022PhRvD.105b3520A}, for HSC-Y1 we use the priors described by \cite{2022PASJ...74..923H}, and for HSC-Y3 we follow \cite{HSCY3-cosmic}. For KiDS-1000, we use correlated priors for the five KiDS tomographic bins \citep{2021A&A...647A.124H}.  When excluding this Gaussian prior, we adopt a wide, flat prior for $\Delta z$ in the range $[-0.2, 0.2]$.  We tested that using a wider prior for $\Delta z$ such as $[-1, 1]$ did not significantly affect our results.

\textbf{Intrinsic Alignment (IA) parameters}: Intrinsic alignment refers to the correlation between the shape of galaxies and large-scale structure not induced by gravitational lensing \citep{2024OJAp....7E..14L}.  In this project, we assume the Non-Linear Alignment (NLA) model \citep{2007Bridle}. The angular cross-power spectrum due to IA can be split into gravitational-intrinsic (GI) and intrinsic-intrinsic (II) parts (see Equation [3-5] in \cite{2007Bridle} and Section 3.2 in \cite{2016MNRAS.456..207K} for details).  The full convergence power spectrum including the effects of intrinsic alignments is,
\begin{equation}
    C^{i,j}_{\mathrm{tot}}(\ell) = C^{i,j}_{\mathrm{GG}}(\ell)+C^{i,j}_{\mathrm{GI}}(\ell) + C^{i,j}_{\mathrm{IG}}(\ell)+C^{i,j}_{\mathrm{II}}(\ell) ,
\label{eq:clkktot}
\end{equation}
where $C^{i,j}_{\mathrm{GG}}$ is the ``gravitational-gravitational'' corresponding to Equation \ref{eq:Limber2}. The Intrinsic-Intrinsic (II) term in Equation \ref{eq:clkktot} can be written as,
\begin{equation}
\label{cii}
C_{\mathrm{II}}^{i,j}(\ell) = \int_0^{\chi_H} \frac{d\chi}{\chi^2} \, n_{\mathrm{s}}^i(\chi) \, n_{\mathrm{s}}^j(\chi) \, P_{\mathrm{II}} \left( k=\frac{\ell + \tfrac{1}{2}}{\chi}, z(\chi) \right) ,
\end{equation}
and the Gravitation-Intrinsic (GI) term can be written as,
\begin{equation}
\label{cgi}
C_{\mathrm{GI}}^{i,j}(\ell) = \int_0^{\chi_H} \frac{d\chi}{\chi^2} \, W_{\mathrm{G}}^i(\chi) \, n_{\mathrm{s}}^j(\chi) \, P_{\mathrm{GI}} \left( k=\frac{\ell + \tfrac{1}{2}}{\chi}, z(\chi) \right),
\end{equation}
where $C_{\mathrm{GI}}^{i,j}$ = $C_{\mathrm{IG}}^{j,i}$.  We note that the gravitation-intrinsic contributions to the galaxy-galaxy lensing model are also scaled by the linear galaxy bias.

For the NLA model, the expression for $P_{\mathrm{II}}$ and $P_{\mathrm{GI}}$ can be written as,
\begin{equation}
    P_{\mathrm{II}}(k,z) = a^2_1(z) \, P_{mm}(k,z) ,
\end{equation}
\begin{equation}
    P_{\mathrm{GI}}(k,z) = a_1(z) \, P_{mm}(k,z) ,
\end{equation}
where,
\begin{equation}
\label{eq_22}
a_1(z) = -A_1 \overline{C}_1 \frac{\rho_\mathrm{crit} \Omega_\mathrm{m}}{D(z)} \left( \frac{1+z}{1+z_0} \right)^{\eta_1} .
\end{equation}
In Equation \ref{eq_22}, $D(z)$ is the linear growth factor and $\overline{C}_1$ is a normalisation constant with value $\overline{C}_1 = 5 \times 10^{-14} \, h^{-2} M_{\odot}^{-1} \text{Mpc}^3$ \citep{2004hirata, 2007Bridle}. The NLA model has two free parameters: the tidal amplitude ($A_1$) and the power-law slope ($\eta_1$).  We use flat priors for these parameters.

\textbf{Multiplicative shear bias parameters}: These calibration parameters are used by each weak lensing survey to obtain the true ellipticity of the sources from the observed shapes.  We denote these parameters as $m_i$, for source bin $i$.  We use Gaussian priors for these parameters as established by the weak lensing surveys, calibrated by image simulations.

\textbf{Lens magnification}: These parameters describe the sign and amplitude of the lens magnification effect \citep[see, e.g.,][]{2023MNRAS.523.3649E}, and are denoted by $\alpha^i$, for lens bin $i$. Mathematically, this effect can be written as,
\begin{equation}
\label{mag}
C^{i,j}_{g\kappa,\mathrm{mag}}(\ell) = 2 \left( \alpha^i - 1 \right) \, C^{i,j}_{mm}(\ell) ,
\end{equation}
where $C^{i,j}_{mm}$ is the convergence power spectrum between the lens and source distribution. This effect contributes to the galaxy-galaxy lensing signal as shown in Equation \ref{eq_14}. For the scale-cut validation, we set the magnification values to zero.  For the shear ratio test, we fix the magnification at the values established by \cite{SvenDESI} for each lens redshift bin, $\alpha^i = [0.94, 1.62, 2.19, 2.54, 2.49, 2.58]$.

\textbf{Baryon feedback}: refers to the influence of the astrophysical processes of star formation, supernovae and active galactic nuclei (AGN) on the distribution of baryons and dark matter in galactic halos, affecting the matter power spectrum on small scales.  On scales $0.1 \, h/\mathrm{Mpc} < k < 10 \, h/\mathrm{Mpc}$, AGNs dominate baryon feedback \citep{bf_scale, bf_scale2}, whilst on scales $k > 10 \, h/\mathrm{Mpc}$, star formation contributes to the effect \citep{bf_scale3}.  We use \textsc{hmcode2020} \citep{hmcode2020} to model baryon feedback, utilizing a single parameterization $\log (T_{\rm AGN}/\mathrm{K})$ corresponding to the AGN heating temperature.  We use a flat prior for $\log (T_{\rm AGN}/\mathrm{K})$ with the range $[7.3, 8.3]$, following \cite{DESblueshear} and model tests from \cite{2024Bigwood}.

We note that we vary different sets of parameters in our validation of the scale cuts (Section \ref{sec:Model_valid}) and in our shear ratio tests (Section \ref{sec:SR_test}).  For the scale-cut validation we vary the galaxy bias ($b$), the baryon feedback parameter ($\log T_{\rm AGN}$), and the cosmological parameters. For the shear ratio test we vary the uncertainty of source redshift ($\Delta z$), the IA parameters ($A_1$ and $\eta_1$), the shear calibration parameters ($m$), and the baryon feedback parameter ($\log T_{\rm AGN}$).  We comment further on these choices in the sections below.  Finally, in our analysis we do not consider the PSF systematic parameters used by the HSC collaboration, deferring a full analysis of these effects to \cite{2025arXiv251215960P}.

\subsection{Likelihood platform}
\label{sec:platform}

When fitting models to the data we perform a Bayesian likelihood analysis.  The Gaussian likelihood of the data, given a model $M$ and set of parameters $\mathbf{p}$, can be written as,
\begin{equation}
    \mathcal{L}(\mathbf{D} | \mathbf{p}, M) \propto e^{-\frac{1}{2} \left[(\mathbf{D} - \mathbf{T}_M(\mathbf{p}))^\mathrm{T} \mathbf{C}^{-1} (\mathbf{D} - \mathbf{T}_M(\mathbf{p}))\right]},
\label{equation_1}
\end{equation}
where $\mathbf{D}$ is the data vector, representing a concatenation of the $3 \times 2$-pt correlation functions $\xi^{i,j}_{\pm}$, $\gamma_{\mathrm{t}}^{i,j}$ and $w_p$, $\mathbf{T}_M(\mathbf{p})$ is the corresponding theoretical model for the given parameters, and $\mathbf{C}$ is the covariance matrix, which is generated as discussed in Section \ref{Covariance}.   We use Bayesian inference to infer the parameters of the theoretical model, given the data, as described by Bayes' theorem,
\begin{equation}
    P(\mathbf{p} | \mathbf{D}, M) \propto \mathcal{L}(\mathbf{D} | \mathbf{p}, M) P(\mathbf{p} | M),
\label{eq_bayes}
\end{equation}
where $P(\mathbf{p} | \mathbf{D}, M)$ is the posterior probability of the model parameters, and $P(\mathbf{p} | M)$ is the prior probability for each parameter.

We sample the posterior probability distribution using the \textsc{nautilus} sampler \citep{2023MNRAS.525.3181L}, which is an algorithm specifically designed for sampling of high-dimensional parameter spaces.  Compared with other nested samplers and MCMC, \textsc{nautilus} has higher sampling efficiency.  For this analysis we configure \textsc{nautilus} with a number of ``live'' points $n_{\mathrm{live}} = 4000$.  We utilise the \textsc{nautilus} sampler within the \textsc{cosmosis} platform \citep{2015A&C....12...45Z}, which provides a modular software framework for cosmological parameter estimation and statistical analysis, and has been widely used for previous weak lensing analyses by survey collaborations.

\section{Model Validation}
\label{sec:Model_valid}

\subsection{Overview}

In order to validate the model and scale cuts for the fits to the $3 \times 2$-pt data vectors combining DESI and weak lensing surveys, we use an analytical method following \cite{2021arXiv210513548K}.  The method is designed to ensure that ``known but unmodelled systematic effects'' do not have a significant impact on the cosmological parameter determinations. We run parameter fits, using the real analysis covariance, to two noise-free data vectors. The first data vector represents a ``\textit{baseline model}'' which corresponds to our fiducial analysis framework, and the second corresponds to a ``\textit{contaminated model}'' which includes additional astrophysical or systematic effects that we do not model in our analysis framework.  We then choose scale cuts such that the difference between the best-fitting values of the key cosmological parameters in these two analyses lies below a small threshold fraction of the statistical errors in these parameters.  Following \cite{2021arXiv210513548K}, we set the threshold as $0.3 \, \sigma$ in the $S_8 - \Omega_{\mathrm{m}}$ 2D plane, such that any systematic effects are significantly sub-dominant to the statistical errors, and potential projection effects that might spuriously affect the test are suppressed.

The ``contaminating'' effects we consider are the modelling of the non-linear matter power spectrum, baryon feedback imprinted in the matter power spectrum, and non-linear galaxy bias.  We consider different values of the scale cut $R_{\rm ggl}$ for galaxy-galaxy lensing ($\gamma_\mathrm{t}(\theta)$, where we convert the projected separation to angles using $\theta = R_{\rm ggl}/\chi(z_{\rm lens})$ with $\chi(z_{\rm lens})$ being the radial comoving distance at the mean redshift of each lens bin) and the scale cut $R_{\rm clus}$ for galaxy clustering ($w_\mathrm{p}(R)$). We performed tests for several different scale cut values, namely $(R_{\rm ggl}, R_{\rm clus}) = (5, 5)$, $(6, 8)$, $(8, 10)$ and $(10, 10) \, h^{-1} \mathrm{Mpc}$.  For the scale cuts on the cosmic shear correlation functions $\xi_+(\theta)$ and $\xi_-(\theta)$, we start from the choices previously adopted by each weak lensing survey as our fiducial setting, and test whether these selections still satisfy our criteria for the expanded $3 \times 2$-pt data vectors including DESI-DR1 data.  For DES-Y3 we follow the scale cuts presented by \cite*{DES_cut}, for HSC-Y1 we use \cite{hsc_cut}, and for HSC-Y3 we adopt the choices of \cite{HSCY3-cosmic}. For KiDS-1000, the fiducial choices of \cite{kids1000_cut2} marginally do not satisfy our criteria, but we find that we can solve this issue using the same scale cuts for $\xi_+(\theta)$, and slightly shifting the $\xi_-(\theta)$ scale cut using the method of \cite*{DES_cut}, as discussed below.

\subsection{Baseline Model}
\label{sec:Baseline}

The baseline model we adopt for the scale cut validation uses the following settings:
\begin{itemize}
\item We use \textsc{hmcode2020} \citep{hmcode2020}, including baryon feedback, to model the non-linear matter power spectrum.  The baryon feedback in this model uses the parameterization $\log T_{\rm AGN}$, where we use the fiducial value $\log T_{\rm AGN} = 7.8$ \citep{bahamas_tagn} in our baseline model.
\item We use a linear galaxy bias model with fiducial values obtained through fits to the projected clustering of the samples in the DESI lens bins presented by \cite{SvenDESI}.
\item We assume the $\Lambda$CDM model and use the DESI fiducial cosmology based on the results of \cite{2020A&A...641A...6P}.
\item We assume the sum of the neutrino masses is $\sum m_\nu = 0.06$ eV (which we also assume for the contaminated data).
\item We set no systematics such as offsets in the source redshift ($\Delta z$), intrinsic alignment, magnification, or multiplicative shear bias.

\end{itemize}
We implement the baseline model in the \textsc{cosmosis} platform, and produce the baseline data using the settings listed above and the correlation models described in Section \ref{sec:models}.

\subsection{Contaminated data}

We produced $3 \times 2$-pt data vectors contaminated with non-linear bias and baryonic effects using hybrid effective field theory (HEFT) \citep{2020MNRAS.492.5754M} as implemented in the \textsc{aemulus $\nu$} emulator \citep{2023JCAP...07..054D}. In particular, we assume the Lagrangian galaxy density field can be expressed as,
\begin{align}
 \label{eq:bias_exp}
    \delta_g[\delta(\bq)] &\approx  1 + b_1 \delta_0 + b_2 (\delta^2_0 - \langle \delta^2_0\rangle) + \\
 & b_\mathrm{s} (s^2_0 - \langle s^2_0 \rangle) + b_{\nabla^2}\nabla^2 \delta_0(\bq) \nonumber \, ,
\end{align}
where the subscript ``$0$'' indicates that a quantity is computed according to the linear field, $s^2_0$ is the square of the traceless tidal tensor, and we have suppressed dependence on the Lagrangian coordinate, $\bq$, in the RHS of this equation. The contribution from $b_{\nabla^2}$ serves two purposes: one as a counter-term in the effective theory and another that captures  short-range non-localities in galaxy formation. Given this Lagrangian field, HEFT replaces the usual perturbative calculations of advected bias operators with a fully non-linear advection operation that makes use of displacements computed from $N$-body simulations:
\begin{align}
    \mo_a(\bx,\tau) &= \int d^3\bq \ \mo_a(\bq)\ \delta_D(\bx - \bq - \Psi(\bq,\tau)) , \nonumber \\
    \mo_a(\bk,\tau) &= \int d^3\bq \ e^{-i\bk\cdot(\bq+\Psi(\bq))}  \ \mo_a(\bq)\,.
    \label{eq:zel_ops}
\end{align}
Given these operators, we can compute galaxy density auto power spectra, $P_{gg}(k)$ and cross-spectra with the matter field, $P_{gm}(k)$,
\begin{align}
    P_{gg}(k) &= \sum_{\mo_i,\mo_j\in \delta_g} b_{\mo_i}b_{\mo_j}P_{\mo_i\mo_j}(k) , \\ 
    P_{gm}(k) &= \sum_{\mo_i\in \delta_g} b_{\mo_j}P_{1\mo_i}(k)\, ,
\end{align}
where $P_{\mo_i\mo_j}(k)$ is the cross-power spectrum of advected operators $\mo_i(k)$ and $\mo_j(k)$, and $b_{\mo_i}$ are bias coefficients corresponding to $\mo_i$. We assume that the bias parameters are different for each sample, and constant with redshift. We set linear bias values for each lens bin based on measured values from the \textsc{Buzzard v2.0} simulations \citep{2024OJAp....7E..57L, 2024arXiv241212548B}, and set non-linear bias values according to the relations in Table 3 of \cite{2022MNRAS.514.5443Z}. When making predictions using a linear bias model we use 
\begin{align}
    P_{gg}(k) &= b_{1,\rm E}^2P^{\rm cdm}_{mm}(k) , \\ 
    P_{gm}(k) &= b_{1,\rm E}P^{\rm cdm}_{mm}(k)\, ,
\end{align}
where $b_{1,\rm E} = b_{\delta}+1$. 

Additionally we express the matter power spectrum in the presence of baryonic effects as
\begin{equation}
    P_{mm}(k) = P^{\rm cdm}_{mm}(k) \left (1 - \frac{b_{\nabla^2 \mathrm{m}} k^2}{1 + (kR)^2}\right) , \,
\end{equation}
motivated by the standard counter-term included in one-loop calculations of the matter power spectrum, and modified with a Padé factor $R = 2\, h^{-1} \textrm{Mpc}$ to control the large-$k$ behavior similar to that used in \cite{2021JCAP...11..026S}. \cite{2024PhRvD.110j3518C} demonstrated the ability of this model to fit a range of baryonic scenarios to $k \sim 2 \, h \, \textrm{Mpc}^{-1}$. We use two options for $P^{\rm cdm}_{\rm mm}(k)$, one making use of the \textsc{aemulus $\nu$} matter power spectrum emulator without including any baryonic feedback effects, and the other using \textsc{hmcode2020} assuming $\log T_{\rm AGN}=7.8$. We additionally consider another case contaminating the \textsc{aemulus $\nu$} matter power spectrum predictions with baryons using $b_{\nabla^2 \rm m}=0.5$,  which leads to a suppression of the matter power spectrum of $\sim10\%$ at $k = 1\, \, \textrm{Mpc}^{-1}$, and well approximates the suppression seen in the \textsc{cosmo-owls} $\log T_{\rm AGN}=8.5$ simulation \cite{2014MNRAS.441.1270L} to $k=2\, h \, \textrm{Mpc}^{-1}$. For the neutrino implementation, in \textsc{aemulus $\nu$} massive neutrinos are included as a separate particle species in the simulations, which interact gravitationally with CDM and baryons. For \textsc{hmcode2020}, the effects of neutrinos on the 1-halo and 2-halo power spectra are described by \cite{hmcode2020}.  Briefly, neutrinos are included in the matter power spectrum which enters the 2-halo term, but do not contribute power to the 1-halo term.  For all these cases, we assume the non-linear galaxy bias prescription described above when generating the contaminated data vectors.  We focus our ensuing analysis on a case that includes all of these contaminating effects, noting that analysis of individual contaminations always produces a lower parameter offset relative to the baseline.

We make predictions for correlation functions given these three dimensional power spectra by computing angular power spectra in the Limber approximation for cosmic shear and galaxy-galaxy lensing, and converting these to the relevant correlation functions using the curved-sky transformations described in \cite{2021arXiv210513548K}. When assuming a linear bias model and \textsc{hmcode2020} matter power spectrum, this pipeline reproduces the predictions of the fiducial pipeline in this work with $\Delta \chi^2\ll 1$.

\subsection{Results}
\label{sec:method_valid}

\begin{table}
\centering
\begin{tabular}{ccc}
\hline
Correlations	&	$(R_{\rm ggl}, R_{\rm clus})$	&	Offset 	\\
\hline
\textbf{KiDS-1000}					\\
$1 \times 2$-pt	&		&	0.293$\sigma$	\\
$2 \times 2$-pt	&	(5, 5) $h^{-1} \mathrm{Mpc}$ 	&	0.176$\sigma$	\\
	&	(6, 8) $h^{-1} \mathrm{Mpc}$	&	0.034$\sigma$	\\
	&	(8, 10) $h^{-1} \mathrm{Mpc}$	&	0.006$\sigma$	\\
	&	(10, 10) $h^{-1} \mathrm{Mpc}$	&	0.011$\sigma$	\\
$3 \times 2$-pt	&	(5, 5) $h^{-1} \mathrm{Mpc}$	&	0.089$\sigma$	\\
	&	(6, 8) $h^{-1} \mathrm{Mpc}$	&	0.155$\sigma$	\\
	&	(8, 10) $h^{-1} \mathrm{Mpc}$	&	0.202$\sigma$	\\
	&	(10, 10) $h^{-1} \mathrm{Mpc}$	&	0.259$\sigma$	\\
\hline				
\textbf{DES-Y3}					\\
$1 \times 2$-pt	&		&	0.178$\sigma$	\\
$2 \times 2$-pt	&	(5, 5) $h^{-1} \mathrm{Mpc}$	&	0.313$\sigma$	\\
	&	(6, 8) $h^{-1} \mathrm{Mpc}$	&	0.025$\sigma$	\\
	&	(8, 10) $h^{-1} \mathrm{Mpc}$	&	0.015$\sigma$	\\
	&	(10, 10) $h^{-1} \mathrm{Mpc}$	&	0.023$\sigma$	\\
$3 \times 2$-pt	&	(5, 5) $h^{-1} \mathrm{Mpc}$	&	0.037$\sigma$	\\
	&	(6, 8) $h^{-1} \mathrm{Mpc}$	&	0.200$\sigma$	\\
	&	(8, 10) $h^{-1} \mathrm{Mpc}$	&	0.196$\sigma$	\\
	&	(10, 10) $h^{-1} \mathrm{Mpc}$	&	0.149$\sigma$	\\
\hline					
\textbf{HSC-Y1}					\\
$1 \times 2$-pt	&		&	0.007$\sigma$	\\
$2 \times 2$-pt	&	(5, 5) $h^{-1} \mathrm{Mpc}$	&	0.317$\sigma$	\\
	&	(6, 8) $h^{-1} \mathrm{Mpc}$	&	0.001$\sigma$	\\
	&	(8, 10) $h^{-1} \mathrm{Mpc}$	&	0.009$\sigma$	\\
	&	(10, 10) $h^{-1} \mathrm{Mpc}$	&	0.023$\sigma$	\\
$3 \times 2$-pt	&	(5, 5) $h^{-1} \mathrm{Mpc}$	&	0.077$\sigma$	\\
	&	(6, 8) $h^{-1} \mathrm{Mpc}$	&	0.009$\sigma$	\\
	&	(8, 10) $h^{-1} \mathrm{Mpc}$	&	0.002$\sigma$	\\
	&	(10, 10) $h^{-1} \mathrm{Mpc}$	&	0.011$\sigma$	\\
\hline					
\textbf{HSC-Y3}					\\
$1 \times 2$-pt	&		&	0.076$\sigma$	\\
$2 \times 2$-pt	&	(5, 5) $h^{-1} \mathrm{Mpc}$	&	0.182$\sigma$	\\
	&	(6, 8) $h^{-1} \mathrm{Mpc}$	&	0.012$\sigma$	\\
	&	(8, 10)	$h^{-1} \mathrm{Mpc}$&	0.009$\sigma$	\\
	&	(10, 10) $h^{-1} \mathrm{Mpc}$	&	0.004$\sigma$	\\
$3 \times 2$-pt	&	(5, 5) $h^{-1} \mathrm{Mpc}$	&	0.214$\sigma$	\\
	&	(6, 8) $h^{-1} \mathrm{Mpc}$	&	0.070$\sigma$	\\
	&	(8, 10) $h^{-1} \mathrm{Mpc}$	&	0.078$\sigma$	\\
	&	(10, 10) $h^{-1} \mathrm{Mpc}$	&	0.185$\sigma$	\\

\hline
\end{tabular}
\caption{The parameter offset between the fits to the baseline and contaminated data vectors in the sub-space of $(\Omega_\mathrm{m}, S_8)$.  We show results corresponding to fits to the $1 \times 2$-pt data (cosmic shear only), $2 \times 2$-pt data (galaxy-galaxy lensing and projected clustering), and to the full $3 \times 2$-pt data vector.  We test scale cuts  $(R_{\rm ggl}, R_{\rm clus}) = (5, 5), \, (6, 8), \, (8, 10)$ and $(10, 10) \, h^{-1} \mathrm{Mpc}$.}
\label{table:valid_diff}
\end{table}

\begin{figure*}
    \centering
    \includegraphics[width=0.33\linewidth]{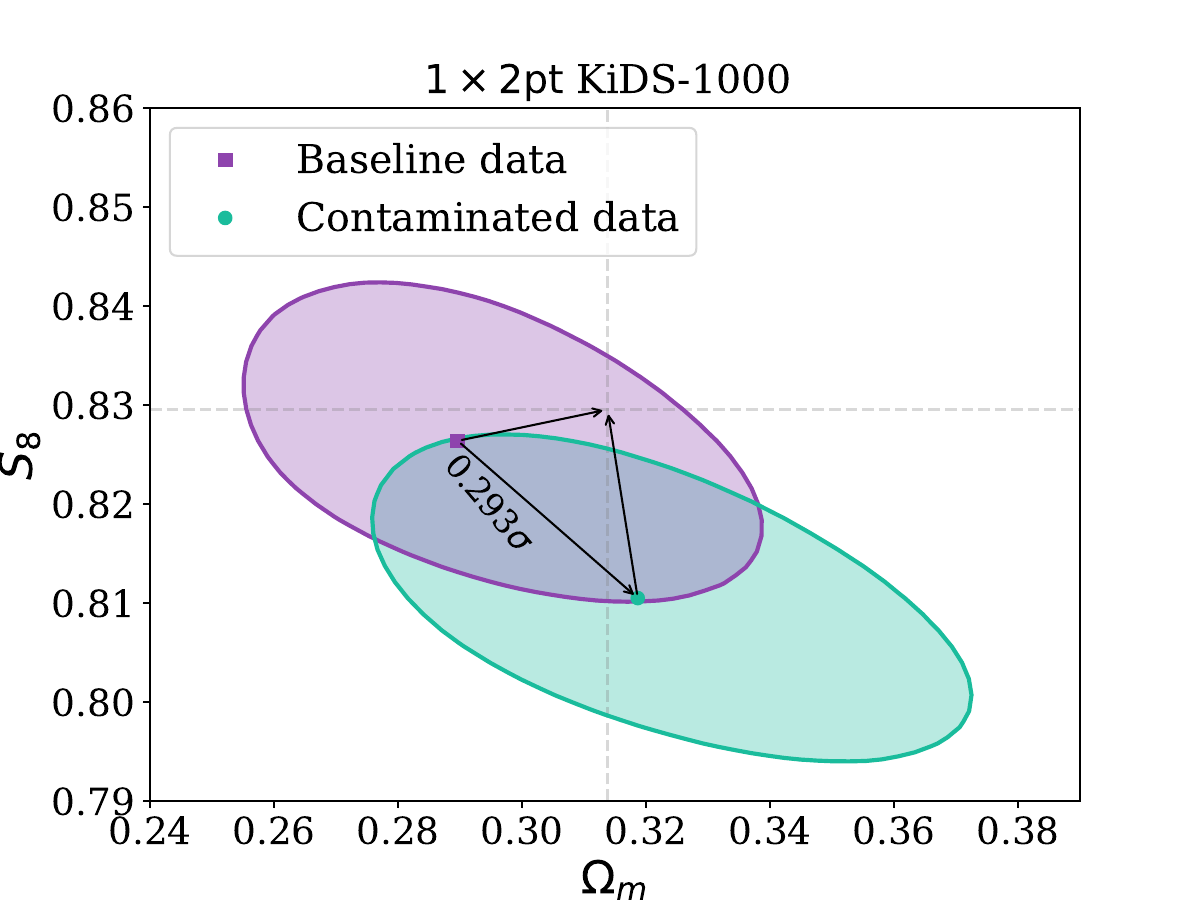}
    \includegraphics[width=0.33\linewidth]{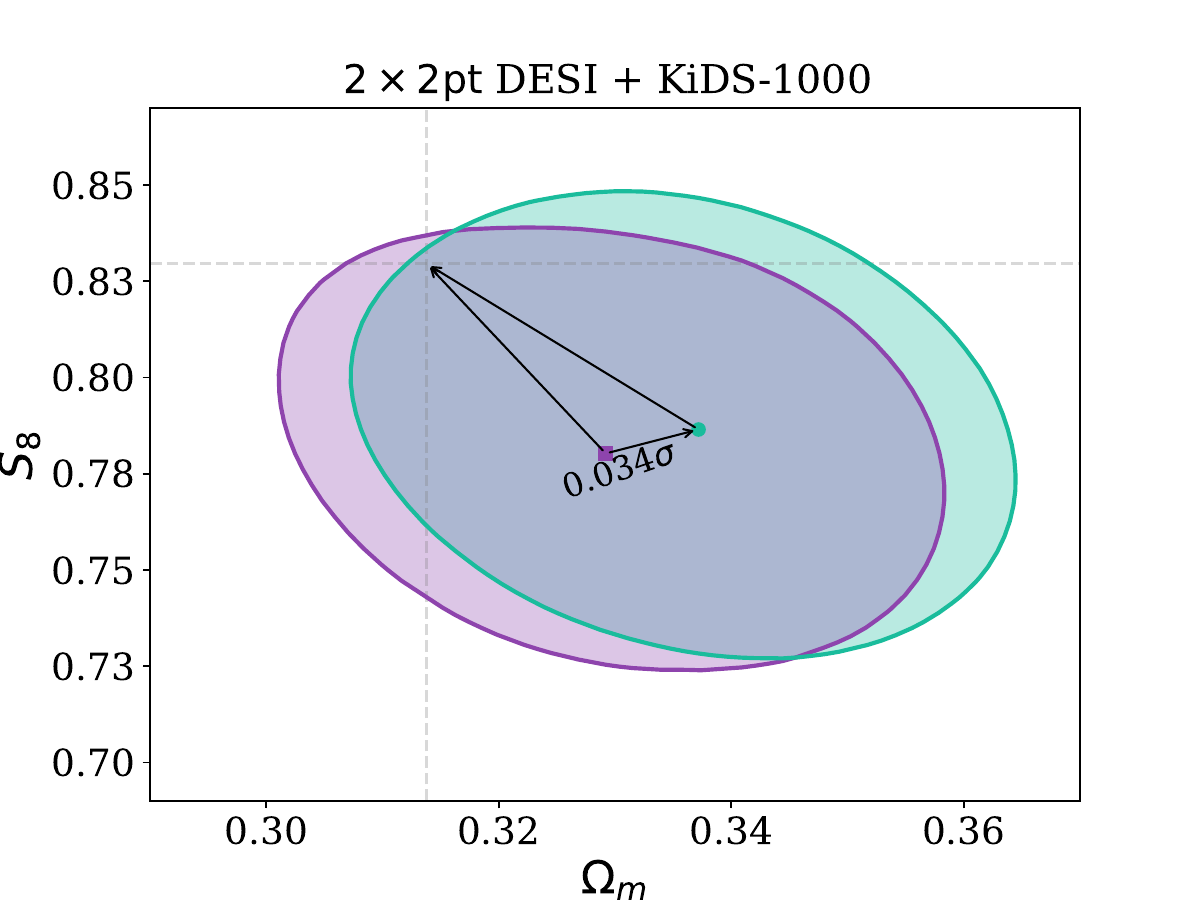}
    \includegraphics[width=0.33\linewidth]{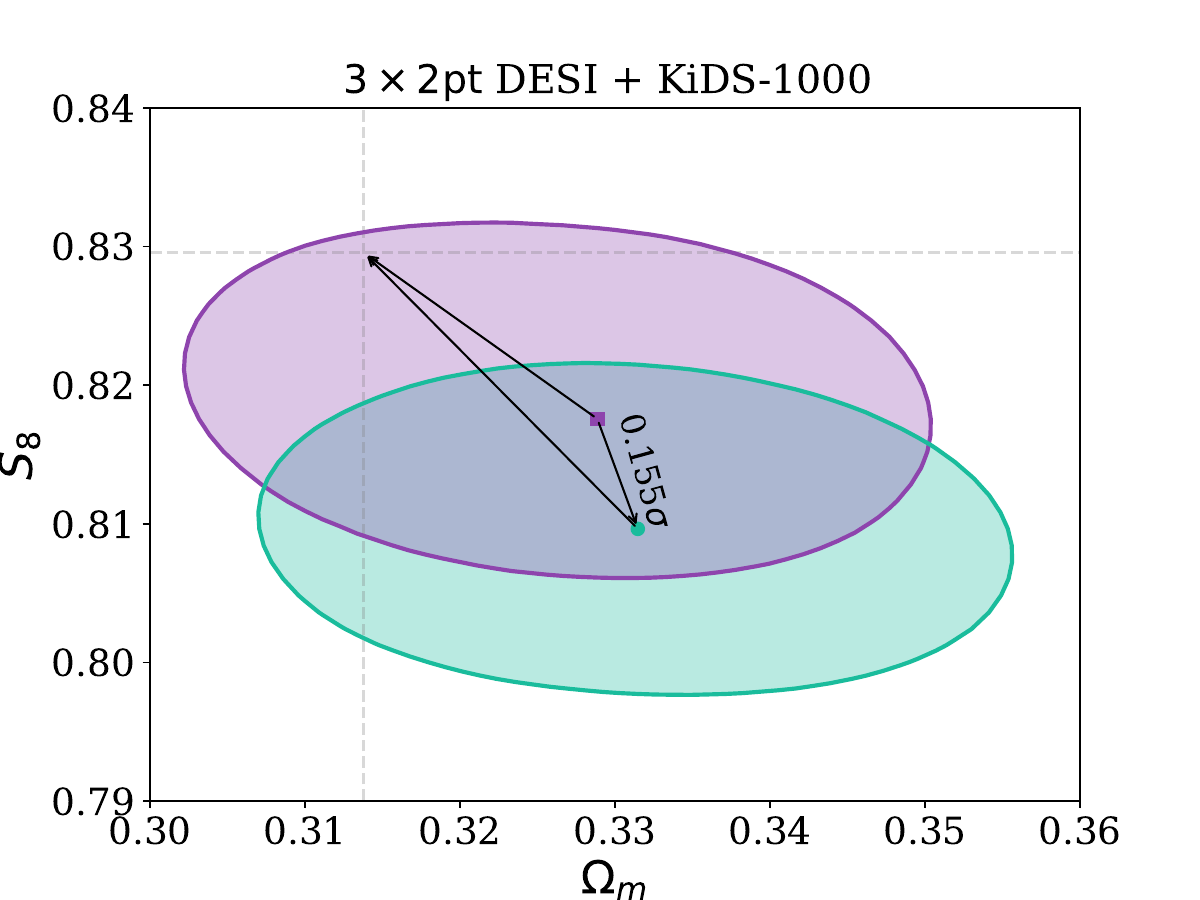}
    \vspace{1\baselineskip}
    \includegraphics[width=0.33\linewidth]{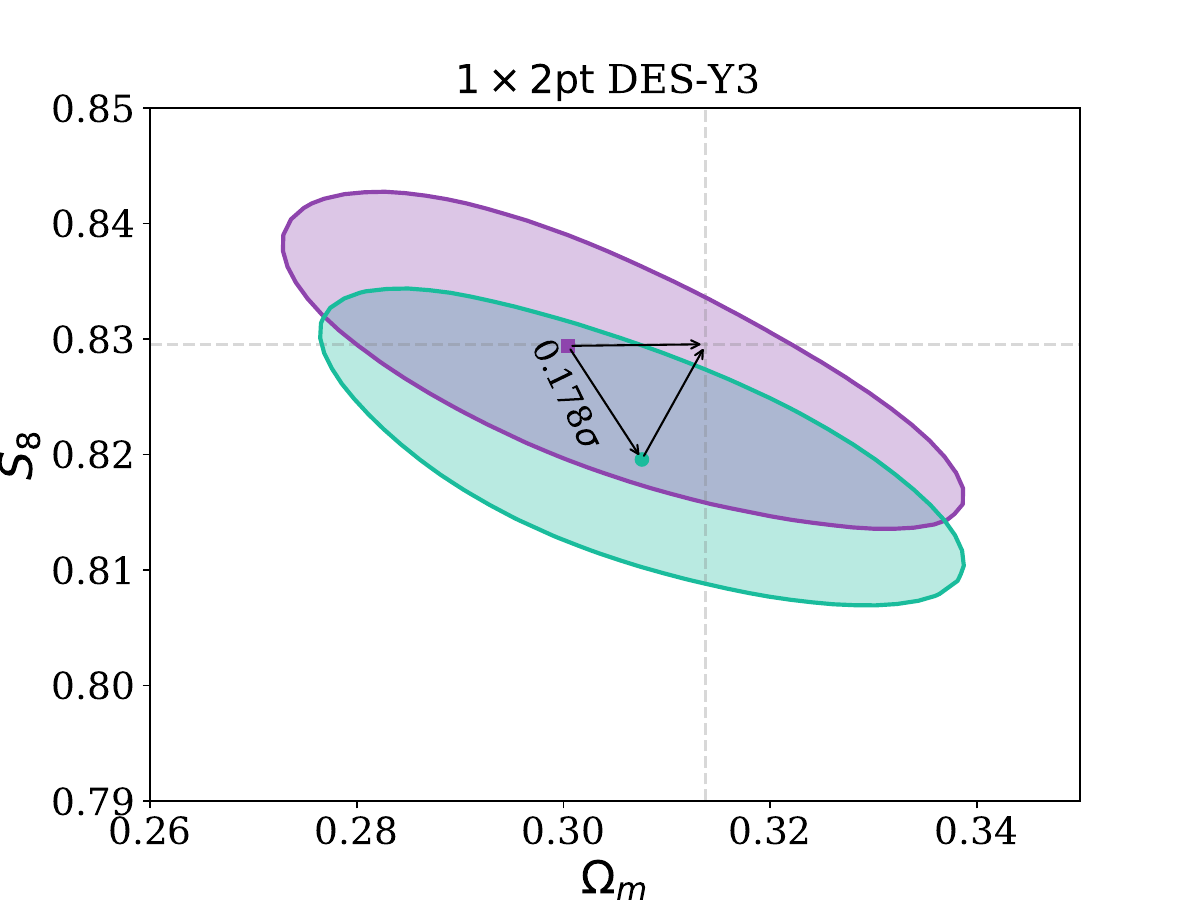}
    \includegraphics[width=0.33\linewidth]{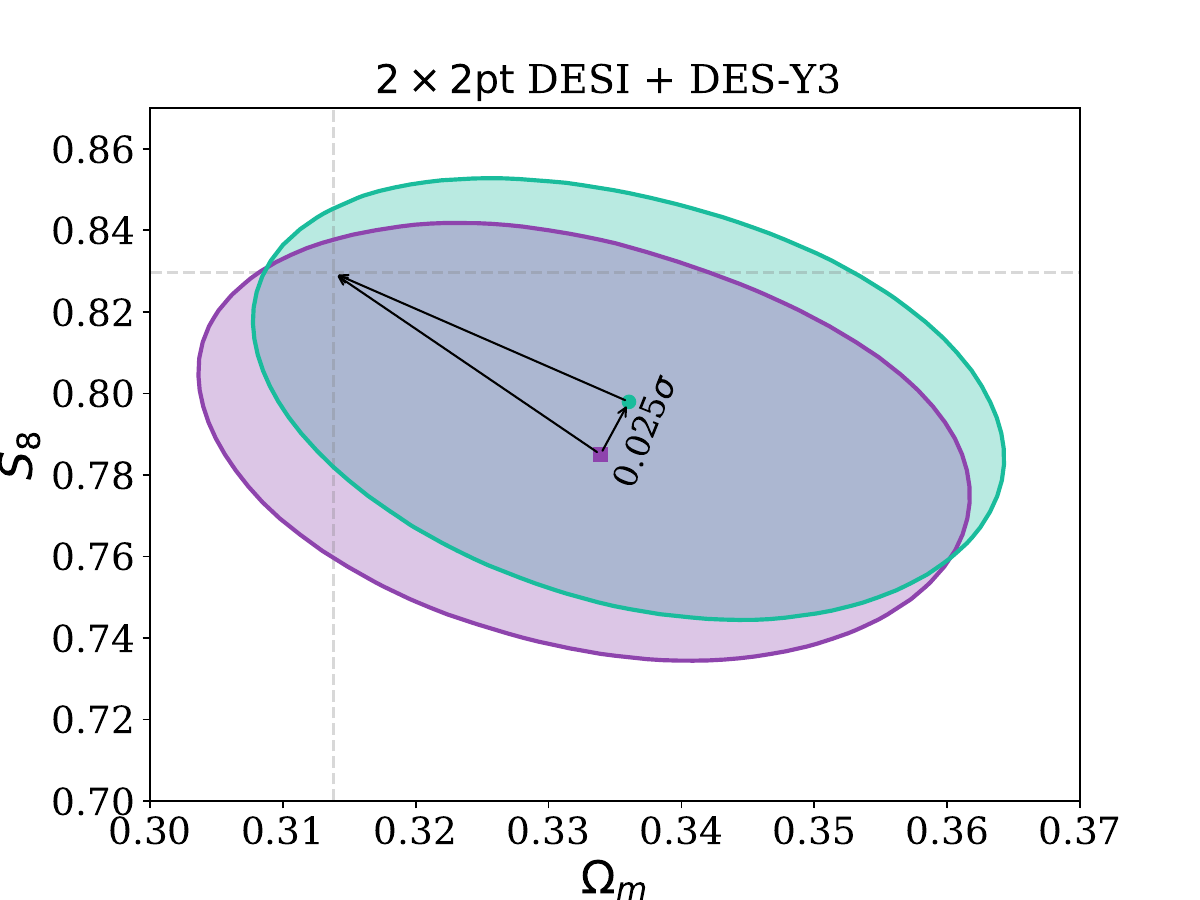}
    \includegraphics[width=0.33\linewidth]{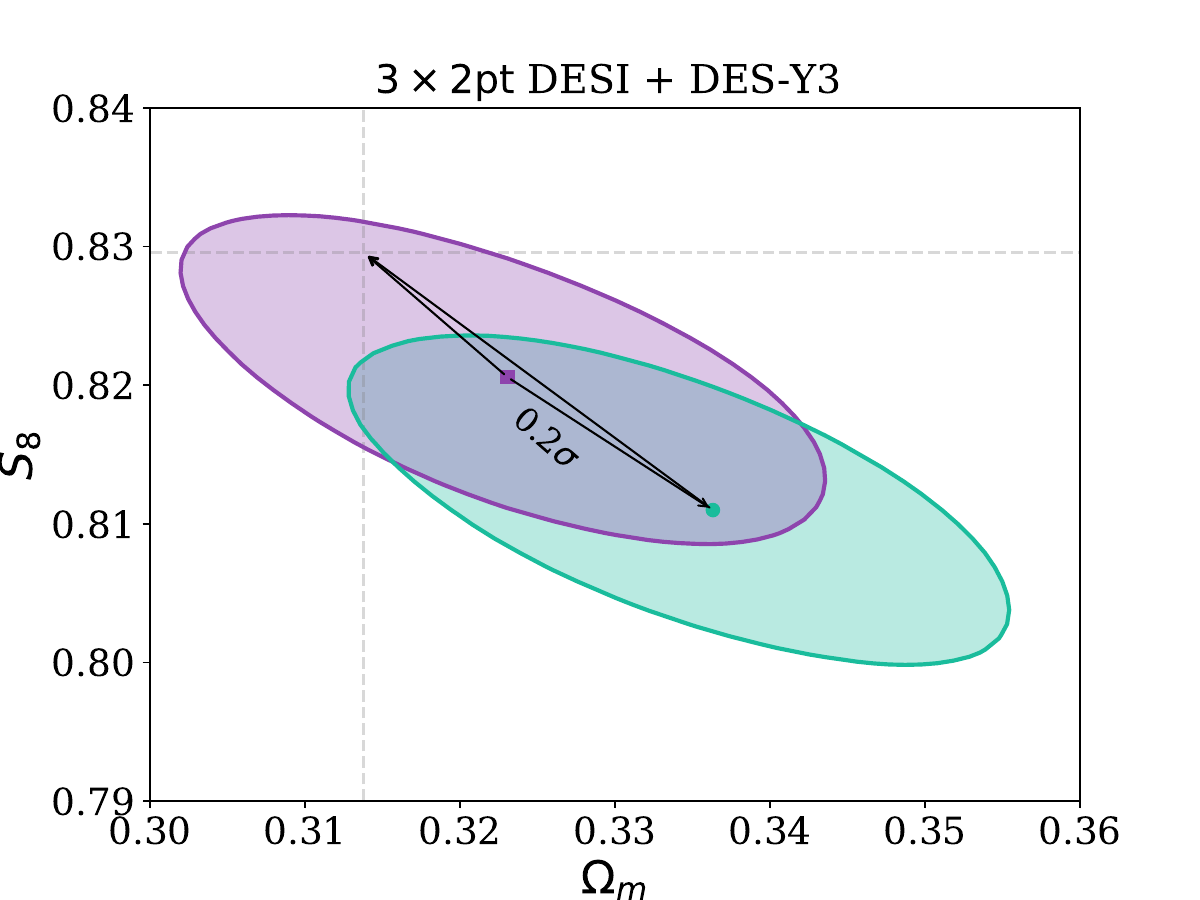}
    \vspace{1\baselineskip}
    \includegraphics[width=0.33\linewidth]{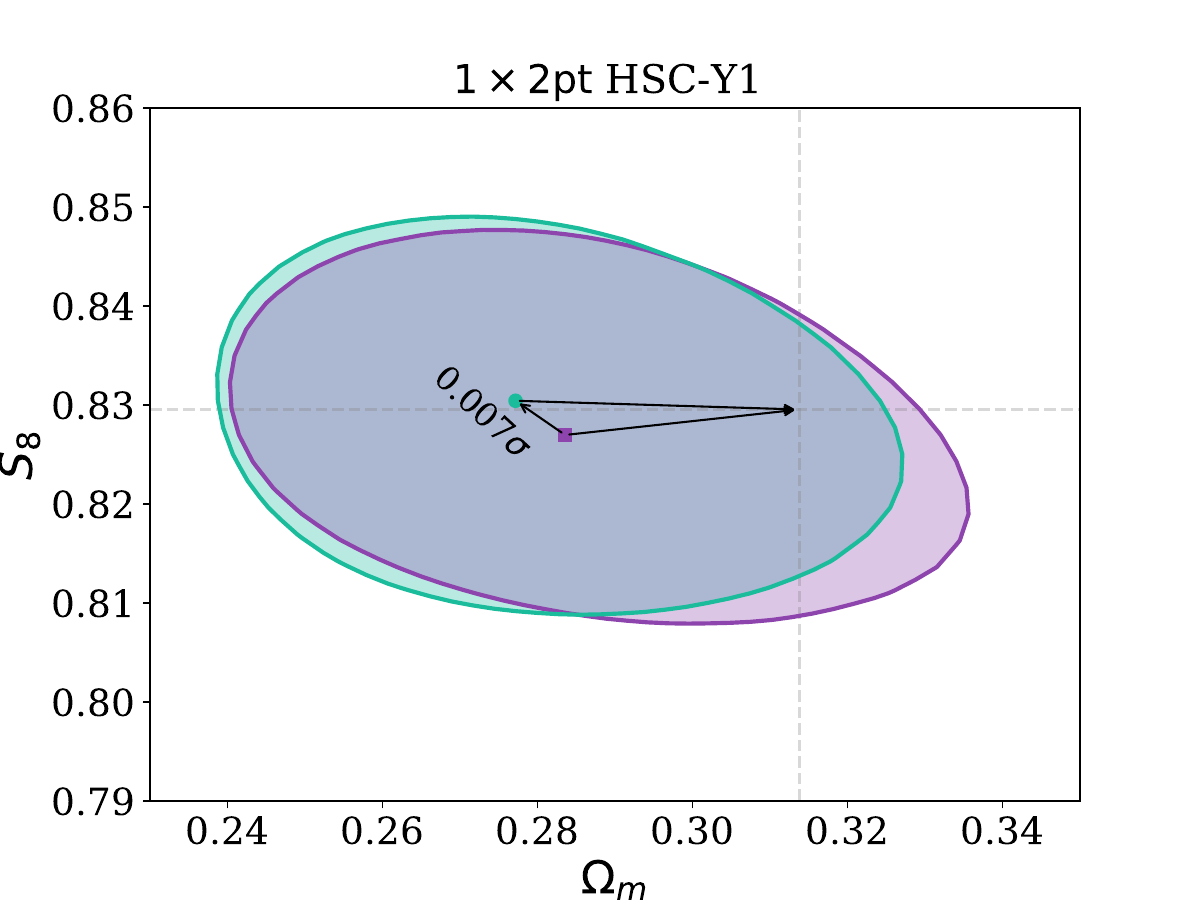}
    \includegraphics[width=0.33\linewidth]{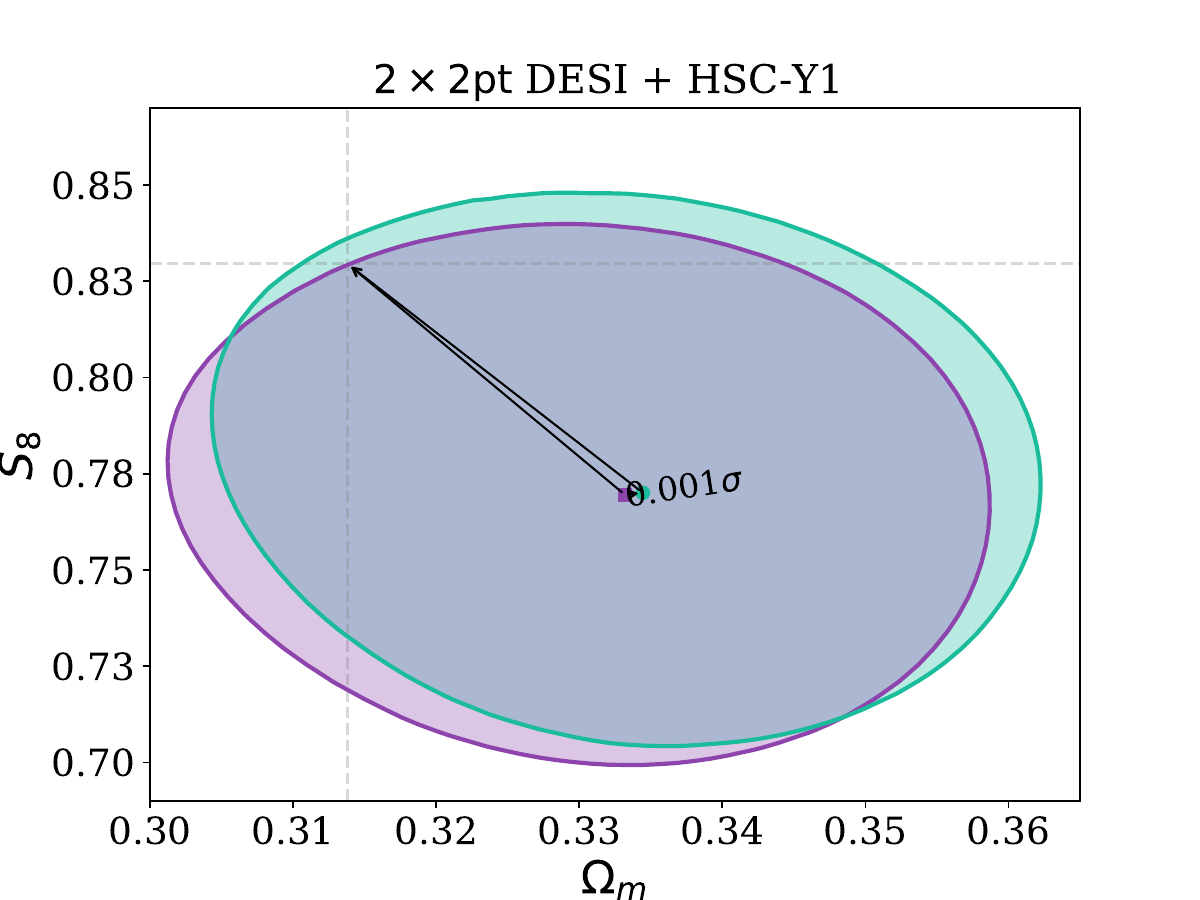}
    \includegraphics[width=0.33\linewidth]{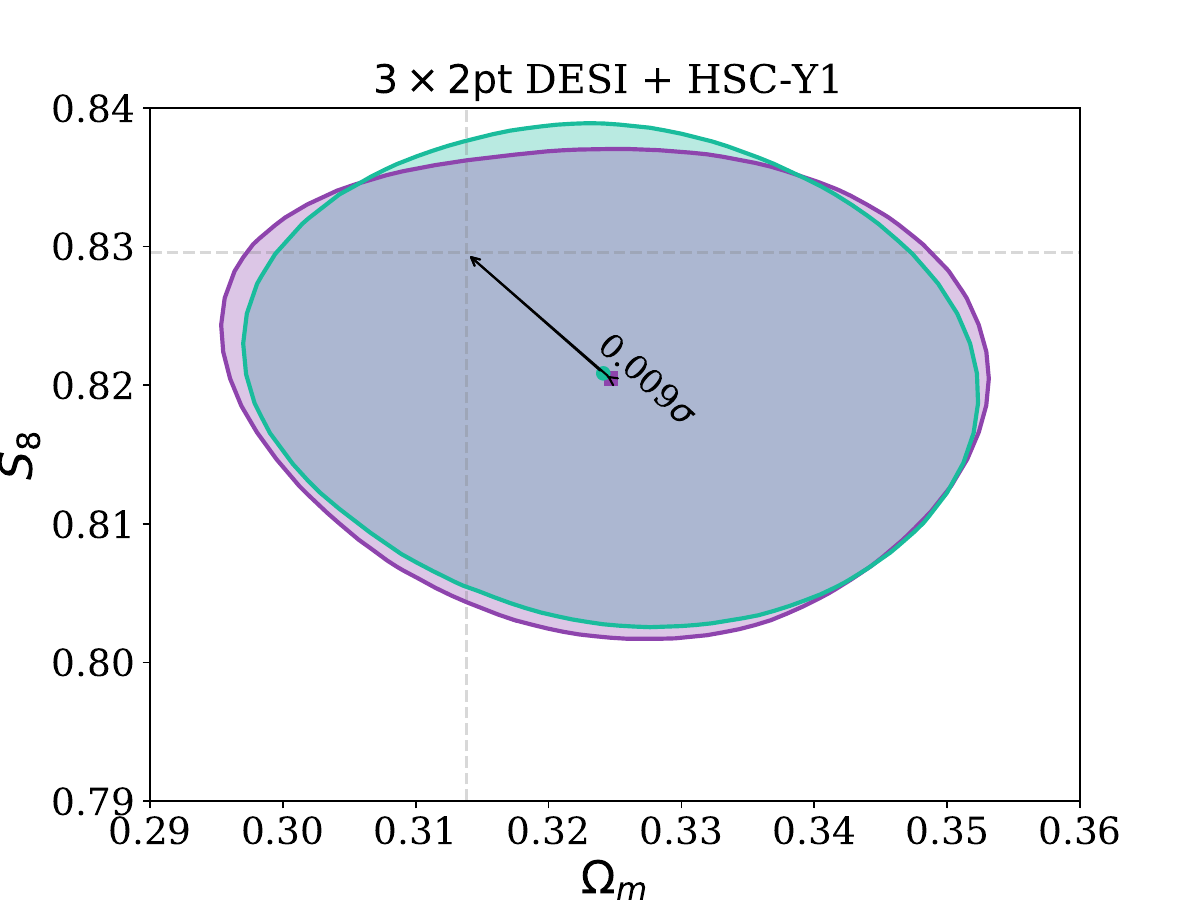}
    \vspace{1\baselineskip}
    \includegraphics[width=0.33\linewidth, trim=0 0 0 0, clip]{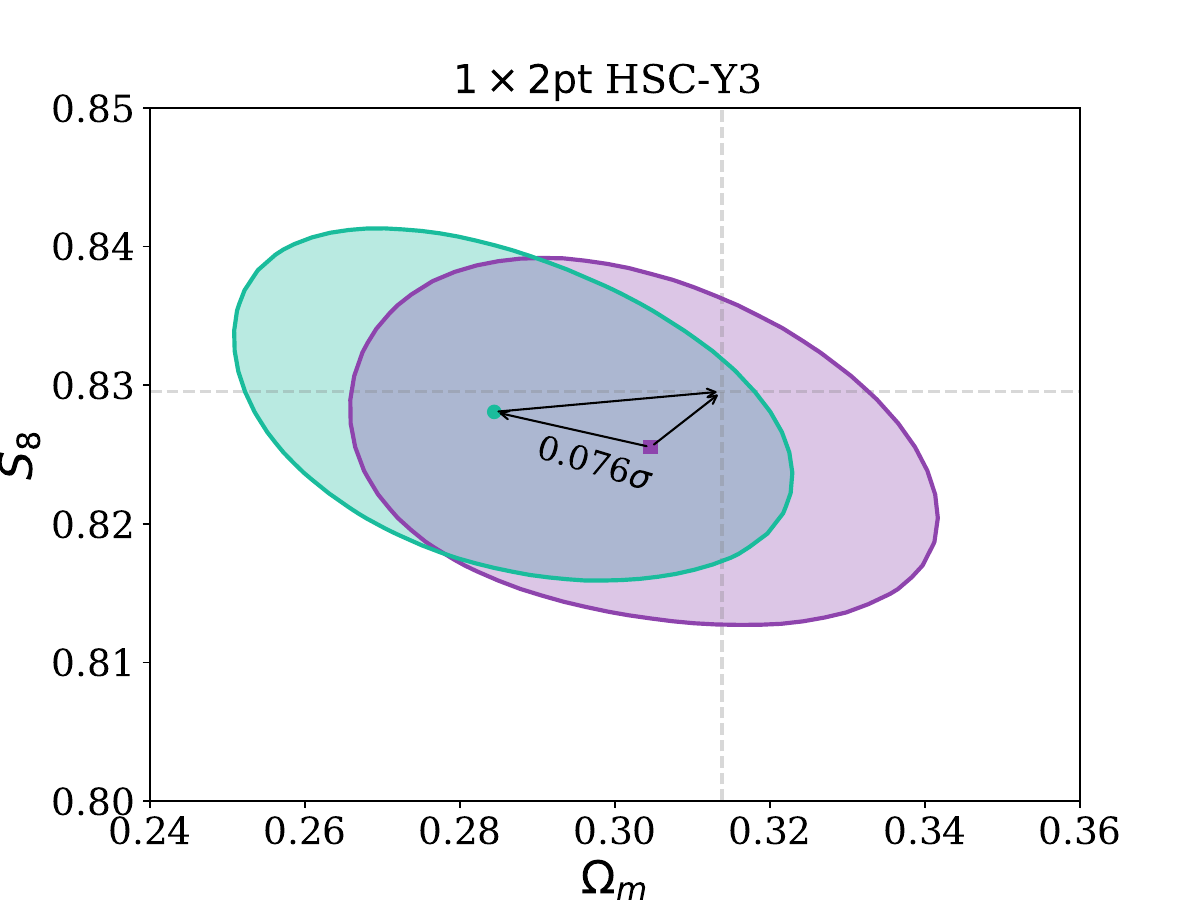}
    \includegraphics[width=0.33\linewidth, trim=0 0 0 0, clip]{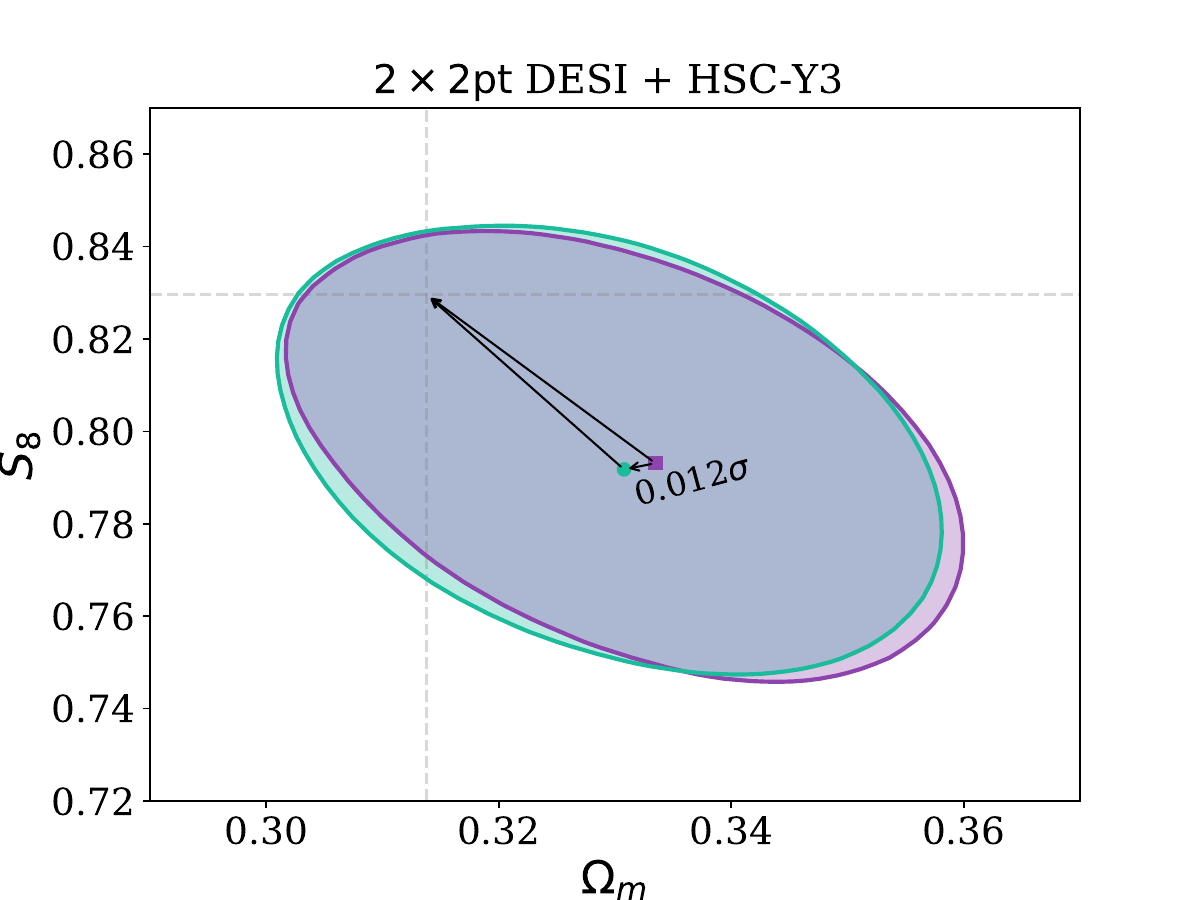}
    \includegraphics[width=0.33\linewidth, trim=0 0 0 0, clip]{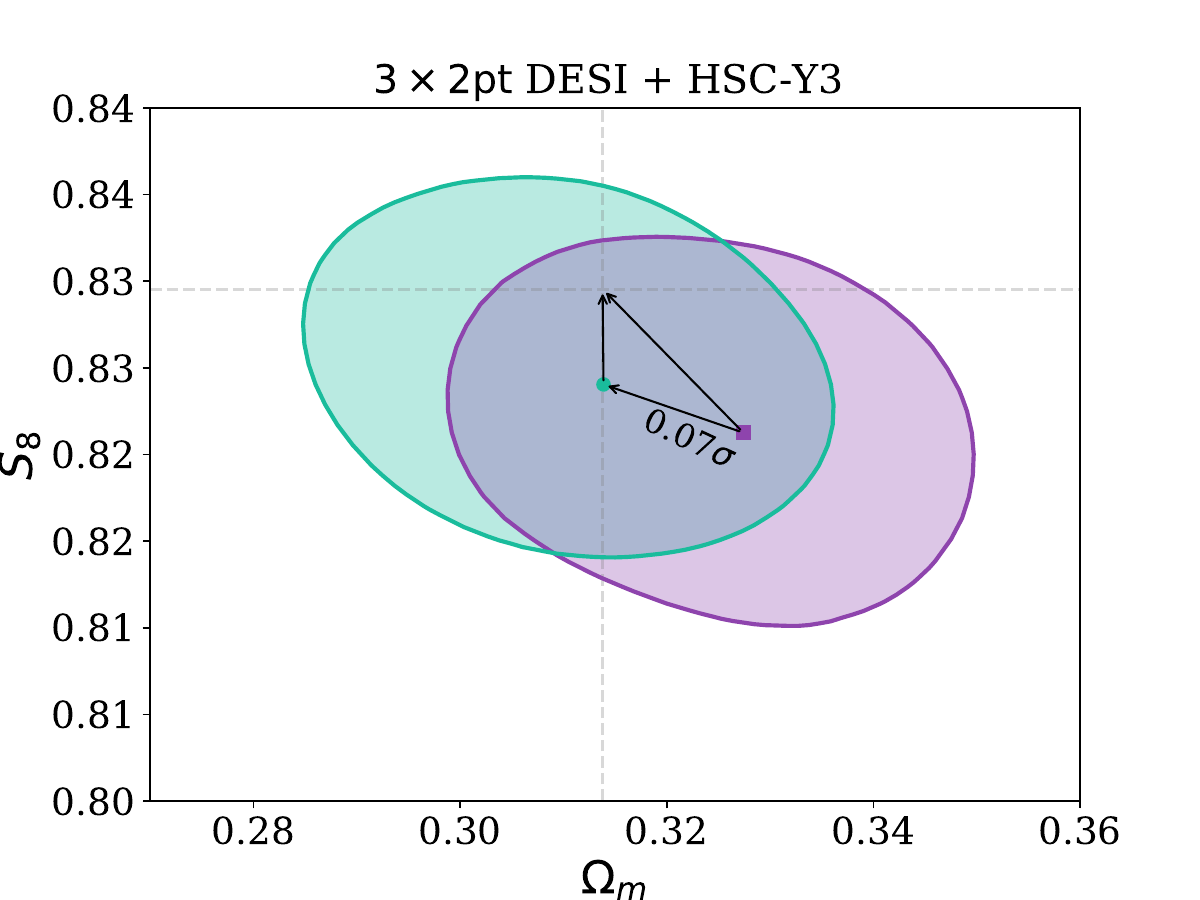}
    \caption{The parameter bias between our baseline and contaminated data vectors assuming a scale cut of $6 \, h^{-1} \, \mathrm{Mpc}$ for galaxy-galaxy lensing and $8 \, h^{-1} \, \mathrm{Mpc}$ for galaxy clustering.  For cosmic shear, we use the fiducial scale cuts of each weak lensing survey, except for a minor update to KiDS-1000 as noted in the Section \ref{sec:method_valid}.  The purple and green shaded regions show the $0.3 \, \sigma$ contour for the 2D marginalized constraints in the sub-space of $(\Omega_\mathrm{m}, S_8)$ for fits to the baseline and contaminated data vectors, respectively.  The different rows show results for KiDS-1000, DES-Y3, HSC-Y1 and HSC-Y3, and the columns display fits to the $1 \times 2$-pt data (cosmic shear only), $2 \times 2$-pt data (galaxy-galaxy lensing and projected clustering), and to the full $3 \times 2$-pt data vector.}
    \label{validation_test}
\end{figure*}

Having created the baseline and contaminated $3 \times 2$-pt noise-free model vectors, we performed a Bayesian likelihood analysis using the real-data covariance for each weak lensing survey.  We use the \textsc{cosmosis} platform \citep{2015A&C....12...45Z} with the \textsc{nautilus} sampler.  We vary the cosmological parameters, linear galaxy bias, and baryon feedback parameters, and we quantify the result of each test using the offset between the fits to the baseline and contaminated models in the projected sub-space of $S_8$ and $\Omega_\mathrm{m}$.  We define the significance of the parameter offset as the \textit{marginalised confidence interval of the fit to the baseline data vector, which intersects the best-fit to the contaminated data vector}.  We repeated these tests for analyses including just cosmic shear (``$1 \times 2$-pt''), galaxy-galaxy lensing and galaxy clustering (``$2 \times 2$-pt'') and all correlations combined (``$3 \times 2$-pt'').

We note that in this test we use a conservative analysis approach in which we set the IA, $\Delta z$ and magnification parameters to zero.  Fixing these systematic parameters reduces the resulting statistical errors in the cosmological parameters, thereby making our $0.3 \, \sigma$ threshold for the allowed variation between the baseline and contaminated datasets more strict.

The offset between the baseline and contaminated data can be seen in Table \ref{table:valid_diff}, for each model survey, combination of correlations, and set of scale cuts $(R_{\rm ggl}, R_{\rm clus})$.  We find that almost all configurations satisfy the  $0.3 \, \sigma$ threshold, with the exception of $(R_{\rm ggl}, R_{\rm clus}) = (5, 5) \, h^{-1} \mathrm{Mpc}$ for the $2 \times 2$-pt correlations of DES-Y3 and HSC-Y1.  In this case, we selected $(R_{\rm ggl}, R_{\rm clus}) = (6, 8) \, h^{-1} \mathrm{Mpc}$ as an acceptable scale cut which passed the threshold requirement for every survey whilst maximizing the separation range used for cosmological analysis. We note that these scale cuts are in agreement with the linear galaxy bias $2 \times 2$-pt scale cuts assumed for the DES-Y3 analysis \citep{desy3_2x2pt}, with which we share the pipeline for galaxy-galaxy lensing.  Figure \ref{validation_test} displays the results of these validation tests for scale cuts $(R_{\rm ggl}, R_{\rm clus}) = (6, 8) \, h^{-1} \mathrm{Mpc}$, for the different weak lensing surveys.

In the case of KiDS-1000 we found that the fiducial choice of cosmic shear scale cuts ($\xi_+< 0.5'$ and $\xi_-< 4'$) marginally did not satisfy the $0.3 \, \sigma$ threshold requirement for our configuration, and hence we explored minor changes.  To select which specific tomographic bins to use for a revised cut, we performed a $\chi^2$ analysis following \cite{2021arXiv210513548K}.  In this method we consider the difference $\Delta\chi^2$ after fitting the baseline model to the contaminated data, individually for the cosmic shear from each pair of source tomographic bins.  The outcome of this analysis was that we could satisfy our $0.3 \, \sigma$  threshold by removing one further data point for $\xi_-$ in bins (2,3), (3,4), (3,5), and (4,5) (and hence modifying the scale cut to $\xi_- < 6.06'$ for those cases).  The results shown in Table \ref{table:valid_diff} assume the new KiDS-1000 cosmic shear scale cut.

The parameter offsets seen in Figure \ref{validation_test} and Table \ref{table:valid_diff} depend on a combination of the fitting range used and the signal-to-noise (S/N) of the different statistics. For $1\times2$-pt (cosmic shear only), the KiDS-1000 offset is larger than that of the other surveys because KiDS-1000 cosmic shear measurements use angular separations down to 0.5 arcmin, which is a smaller scale than the other surveys, and the contaminated model offset compared to the baseline increases at smaller scales. For the GGL and galaxy clustering offsets ($2\times2$-pt), the offsets are comparable because we assume identical fitting ranges for different lensing surveys.  In this case, the remaining offsets are driven by variations in S/N and parameter degeneracies. Since our results are based on an MCMC analysis with finite chains, the estimated offsets are also subject to statistical noise, which we estimate to be approximately 0.1$\sigma$.

Finally, we repeated these scale cut validations for a $w$CDM model, considering both the $S_8 - \Omega_\mathrm{m}$ and $w - \Omega_\mathrm{m}$ sub-spaces.  We found that the same scale cuts as above, $(R_{\rm ggl}, R_{\rm clus}) = (6, 8) \, h^{-1} \mathrm{Mpc}$ with our amended scale cut for KiDS-1000, satisfied the $0.3 \, \sigma$ threshold requirement in the parameter offset for the $3 \times 2$-pt statistics.

\section{Shear Ratio Tests}
\label{sec:SR_test}

\subsection{Shear ratio measurements}
\label{sec:SR_meas}

\begin{figure*}
    \centering
    \includegraphics[width=0.49\linewidth]{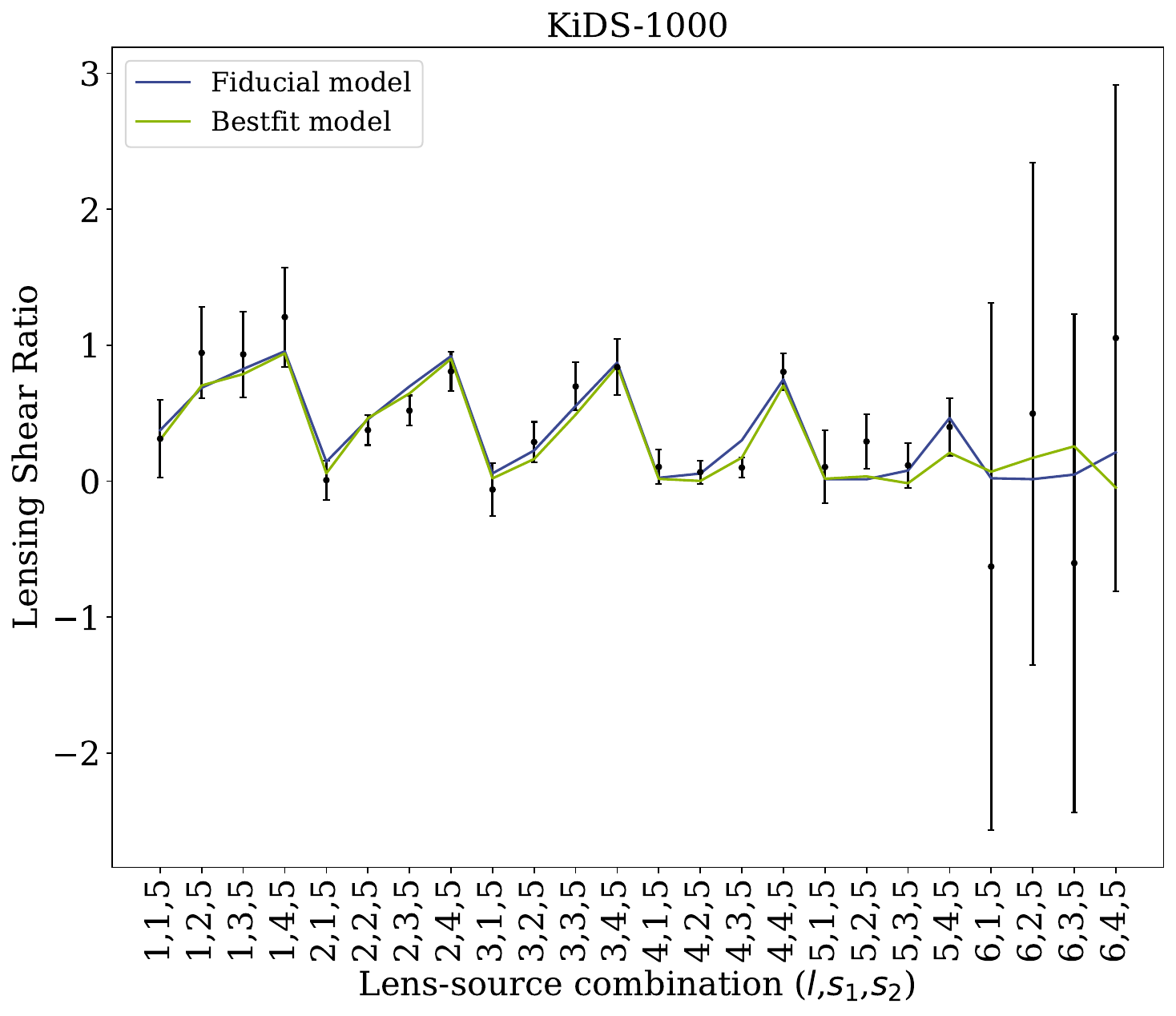}
    \includegraphics[width=0.49\linewidth]{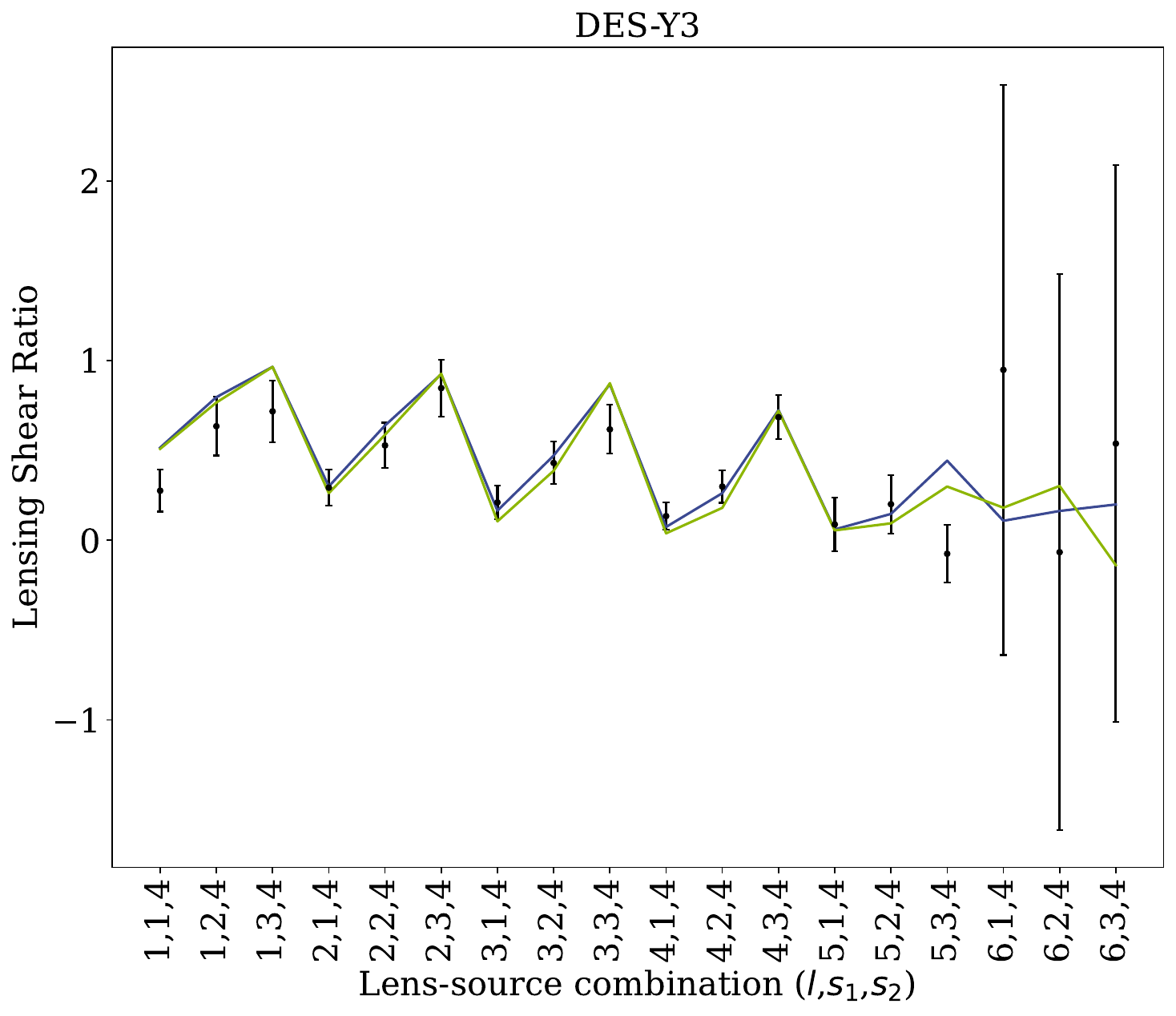}
    \vspace{1\baselineskip}
    \includegraphics[width=0.49\linewidth]{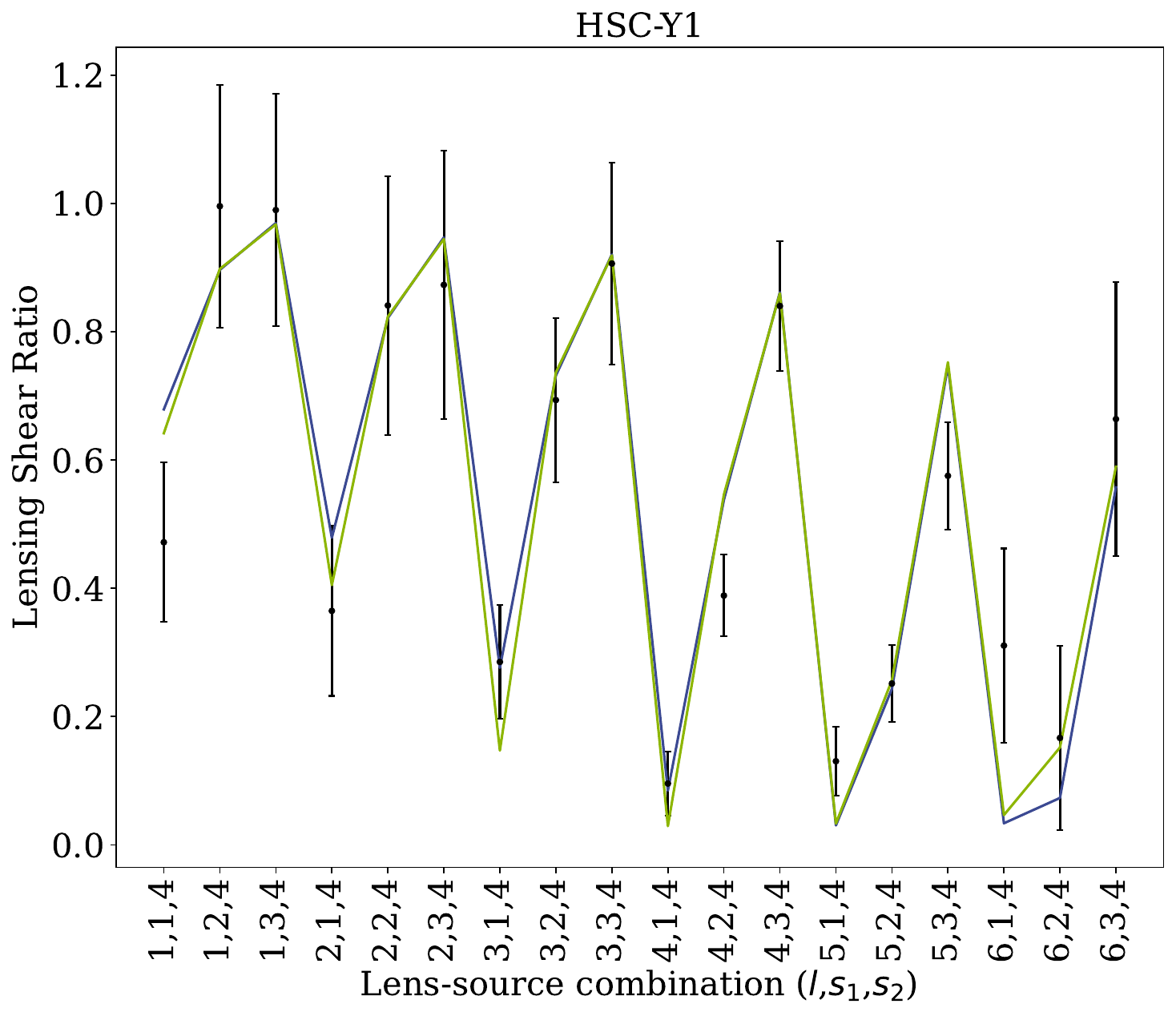}
    \includegraphics[width=0.49\linewidth]{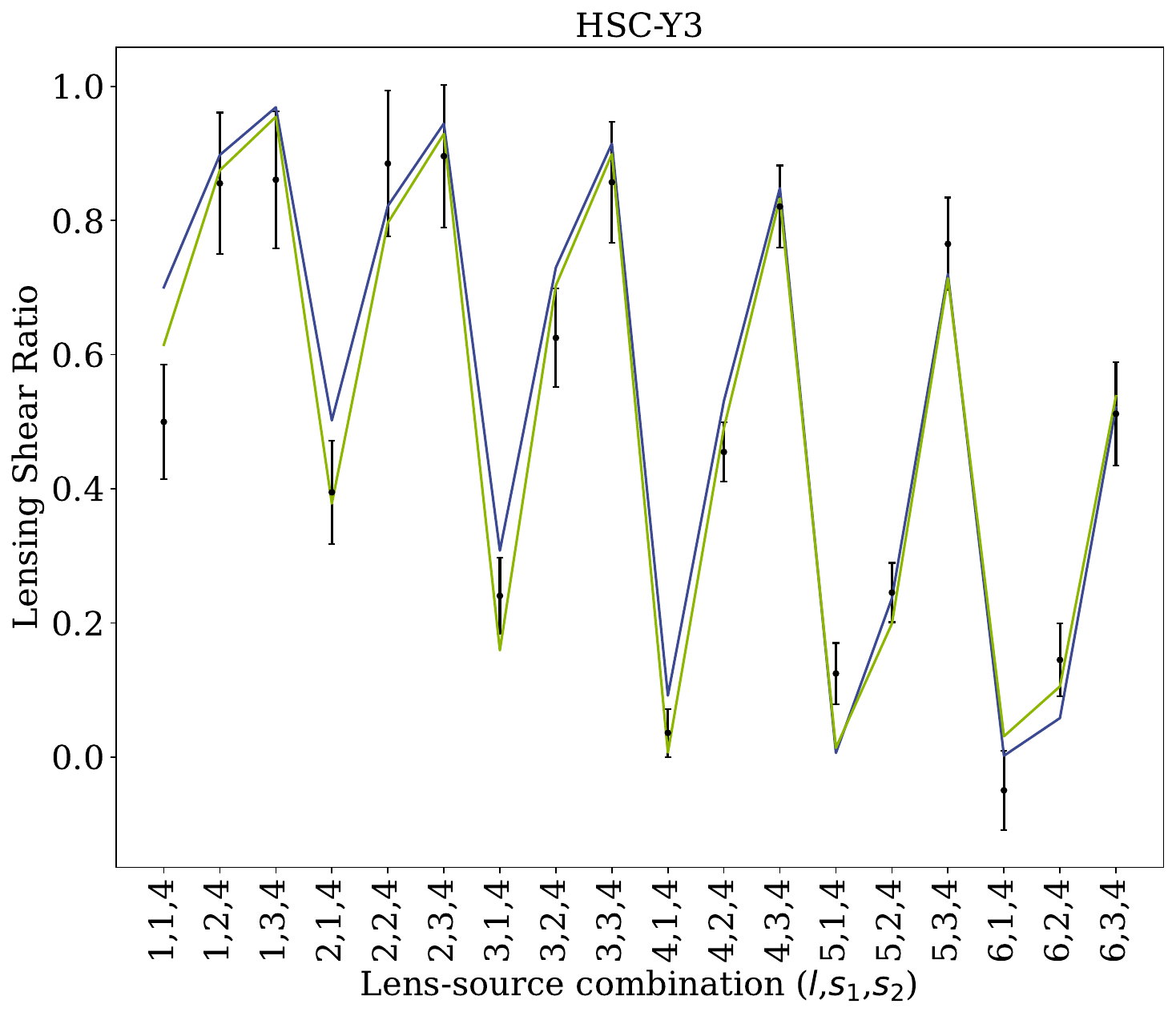}
    \caption{The measurements of the angle-averaged shear ratio from DESI-DR1 lenses and KiDS, DES-Y3, HSC-Y1 and HSC-Y3 sources.  The averaged ratios are displayed for all the unique combinations of lens and source bins, ordered in groups corresponding to a given lens sample.  We overplot the best-fitting model (green solid line) from our baseline analysis, and additionally a model in which we set the values of IA, $\Delta z$, $m$ and magnification equal to zero (the blue solid line, which we refer to as the fiducial model).}
    \label{Fid_best_SR}
\end{figure*}

The shear ratio is a statistic that compares the average tangential shear of two different source samples around the same lens sample, at a given angular separation $\theta$.  It is defined by the equation,
\begin{equation}\label{eq_9}
r^{(\mathrm{l,s_i,s_j})}(\theta) = \frac{\gamma^{\mathrm{l,s_i}}_\mathrm{t}(\theta)}{\gamma^{\mathrm{l,s_j}}_t(\theta)} = \frac{\Sigma_\mathrm{crit}^{-1}(z_\mathrm{l},z_\mathrm{si})}{\Sigma_\mathrm{crit}^{-1}(z_\mathrm{l},z_\mathrm{sj})} ,
\end{equation}
where $\mathrm{l}$ denotes the lens sample and $\mathrm{s_i}$ and $\mathrm{s_j}$ denote two different source samples.  The second equality in Equation \ref{eq_9} applies in the case of an individual source-lens pair, where it can be seen that the shear ratio traces geometrical distances through the $\Sigma_\mathrm{crit}$ parameter.  In the case of lens and source tomographic samples, we must integrate over their redshift distributions when modelling the average tangential shear.  We always use the highest source tomographic bin in the denominator of Equation \ref{eq_9}, following \citet*{2022PhRvD.105h3529S}, and consider the remaining source tomographic bins in the numerator.  We use the highest tomographic bin as the denominator since it generally has the highest signal-to-noise of the measurements.

As already discussed in Section \ref{sec:methods}, we adopt the model for the average tangential shear introduced in \citet*{2022PhRvD.105h3529S} that includes astrophysical effects such as intrinsic alignment, uncertainty of source redshift, magnification and multiplicative shear bias parameters.  This allows us to constrain a range of astrophysical parameters using the shear ratio.

To an excellent approximation, the shear ratio is independent of the source-lens angular separation $\theta$ \citep[see Equation 6 and 7 in][]{Emas_SR}.  Therefore we can compress the data by averaging the shear ratio $r^{(\mathrm{l,s_i,s_j})}(\theta)$ over a range of angular separations,
\begin{equation}
r^{(\mathrm{l,s_i,s_j})} \equiv \left \langle \frac{\gamma^{\mathrm{l,s_i}}_\mathrm{t}(\theta)}{\gamma^{\mathrm{l,s_j}}_\mathrm{t}(\theta)} \right \rangle_\theta = \left \langle r^{(\mathrm{l,s_i,s_j})}(\theta) \right \rangle_\theta ,
\label{eq:averagesr}
\end{equation}
and we express our final shear ratio measurement in this angle-averaged form for each combination of $(\mathrm{l,s_i,s_j})$.

An estimator based on a ratio, such as Equation \ref{eq_9}, is subject to bias if the denominator is noisy or close to zero \citep{bayes_fit}, which can occur when a significant portion of the source distribution is located in front of the lenses, for example for high-redshift lens distributions.  Therefore, we estimate shear ratios and their posterior probability distributions not by a direct ratio, but rather by using an unbiased Bayesian maximum-likelihood method, where the application to the shear ratio has been described in detail by \cite{Emas_SR}.  The measurement of the angle-averaged shear ratio using the Bayesian method for DESI-DR1 data and the weak lensing surveys considered in this study is shown in Figure \ref{Fid_best_SR}.  In this figure we display the angle-averaged ratios for all the unique combinations of lens and source bins, ordered in groups corresponding to a given lens sample.

\subsection{Methodology of SR validation}

In this section we fit astrophysical models to a combination of the DESI-DR1 shear ratios and the ``$1 \times 2$-pt'' shear correlation function datasets $[\xi_+(\theta), \xi_-(\theta)]$ for KiDS-1000, DES-Y3, HSC-Y1 and HSC-Y3 publicly released by the weak lensing collaborations.  The capacity of a consistent set of model parameters to describe both the cosmic shear and SR measurements serves as an additional validation of our modelling choices.  In the scope of this test, we fix the values of the cosmological parameters at the fiducial values of the DESI collaboration, which follow the Planck 2018 TT,TE,EE+lowE+lensing best-fitting cosmology \citep{2020A&A...641A...6P}.  We note that cosmological parameter fits (after data-vector blinding) will be presented by \cite{2025arXiv251215960P} for the full $3 \times 2$-pt correlations.

We perform a Bayesian likelihood analysis to fit the model parameters to our datasets, using the \textsc{cosmosis} platform \citep{2015A&C....12...45Z} with the \textsc{nautilus} sampler.  We fix the cosmological, galaxy bias and magnification parameters.  The cosmic shear and shear ratio models are independent of galaxy bias (in the linear bias model we are adopting here), and the magnification parameters have been measured for our DESI lens bins by \cite{SvenDESI}.  We vary the intrinsic alignment parameters in the Non-Linear Alignment (NLA) model $(A_1, \eta_1)$, the $\Delta z_i$ parameters describing the offset of the mean redshift of the source redshift distributions for each tomographic bin, and the multiplicative shear calibration bias parameters $m_i$ for each source tomographic bin, where for the latter we use the priors established by each weak lensing collaboration.  The parameter ranges and priors can be seen in Table \ref{table:priors}.

In our baseline analysis we average the shear ratio measurements over an angular range corresponding to projected separation $[2-6] \, h^{-1}$ Mpc at the lens redshifts, following \citet*{2022PhRvD.105h3529S}.  The small-scale cut-off in their analysis was motivated by caution regarding extending the fit into the ``1-halo'' regime, as tested using simulations.  The large-scale cut-off was motivated by not double-counting information, given that $6 \, h^{-1}$ Mpc was the minimum cut-off for including $\gamma_\mathrm{t}$ data in the $3 \times 2$-pt analysis of \citet*{2022PhRvD.105h3529S}.  In our analysis, we do not separately fit to $\gamma_\mathrm{t}$ data and only combine the shear ratio with cosmic shear information, so we can average the shear ratio measurements over a wider range of scales without double-counting.  Hence, we also consider using an extended separation range $[1-10] \, h^{-1}$ Mpc as well as an unrestricted average over all separations.  Sec 5.3 of \cite{Emas_SR} demonstrated that information from the shear ratio can be somewhat increased by expanding the range of scales, improving the parameter constraints.

In our baseline analysis we model the non-linear power spectrum using \textsc{hmcode2020} \citep{hmcode2020} with baryon feedback.  However, we also test the sensitivity of our results to different modelling choices: \textsc{hmcode2020} without baryon feedback, and \textsc{halofit} \citep{Takahashi_2012}.  Finally, we test the sensitivity of our results to changing the fiducial cosmological parameters from the DESI cosmology (with $\Omega_\mathrm{m} = 0.3153$) to the best-fitting parameters determined by each weak lensing survey.  For KiDS-1000, we adopt the values described in \cite{kids1000_cut2}; for DES-Y3, we follow \cite{2022PhRvD.105b3514A}; for HSC-Y1, we rely on \cite{2019PASJ...71...43H}; and for HSC-Y3, we use the value in \cite{HSCY3-cosmic}.  These values are $\Omega_\mathrm{m} = [0.246, 0.289, 0.365, 0.256]$ and $S_8 = [0.759, 0.772, 0.800, 0.769]$ for KiDS-1000, DES-Y3, HSC-Y1 and HSC-Y3, respectively.  In our analysis we specifically focus on the constraints on the $\Delta z$ and IA parameters enabled by the shear ratio, using the same flat priors for every lensing survey.

\subsection{Results}

\subsubsection{Baseline parameter fits}

\begin{table*}
\centering
\begin{tabular}{cccccc}
\hline
Lensing survey & Parameter & Improvement & Improvement & Shift & Shift \\															
& &  without $\Delta z$ prior & with $\Delta z$ prior &without $\Delta z$ prior & with $\Delta z$ prior  \\															
\hline															
KiDS-1000	&	$\Delta z_1$	&	-11.36	\%	&	-2.69	\%	&	0.77	$\sigma$	&	0.07	$\sigma$	\\
	&	$\Delta z_2$	&	-1.12	\%	&	2.25	\%	&	0.78	$\sigma$	&	0.14	$\sigma$	\\
	&	$\Delta z_3$	&	-0.65	\%	&	4.65	\%	&	0.60	$\sigma$	&	0.31	$\sigma$	\\
	&	$\Delta z_4$	&	-2.07	\%	&	1.14	\%	&	0.25	$\sigma$	&	0.13	$\sigma$	\\
	&	$\Delta z_5$	&	-4.89	\%	&	0.22	\%	&	0.03	$\sigma$	&	0.04	$\sigma$	\\
	&	$A_1$	&	36.41	\%	&	26.90	\%	&	1.40	$\sigma$	&	0.54	$\sigma$	\\
\hline															
															
DES-Y3	&	$\Delta z_1$	&	29.56	\%	&	1.67	\%	&	0.42	$\sigma$	&	0.25	$\sigma$	\\
	&	$\Delta z_2$	&	30.96	\%	&	2.91	\%	&	0.47	$\sigma$	&	0.07	$\sigma$	\\
	&	$\Delta z_3$	&	33.87	\%	&	1.44	\%	&	0.03	$\sigma$	&	0.44	$\sigma$	\\
	&	$\Delta z_4$	&	40.94	\%	&	1.96	\%	&	0.69	$\sigma$	&	0.25	$\sigma$	\\
	&	$A_1$	&	44.24	\%	&	13.77	\%	&	0.63	$\sigma$	&	0.08	$\sigma$	\\
\hline															
															
HSC-Y1	&	$\Delta z_1$	&	32.72	\%	&	2.57	\%	&	0.45	$\sigma$	&	0.32	$\sigma$	\\
	&	$\Delta z_2$	&	29.98	\%	&	2.87	\%	&	1.36	$\sigma$	&	0.68	$\sigma$	\\
	&	$\Delta z_3$	&	23.84	\%	&	1.29	\%	&	0.76	$\sigma$	&	0.15	$\sigma$	\\
	&	$\Delta z_4$	&	12.85	\%	&	0.09	\%	&	0.52	$\sigma$	&	0.32	$\sigma$	\\
	&	$A_1$	&	66.03	\%	&	34.47	\%	&	0.62	$\sigma$	&	1.36	$\sigma$	\\
\hline															
															
HSC-Y3	&	$\Delta z_1$	&	48.90	\%	&	0.51	\%	&	0.51	$\sigma$	&	0.91	$\sigma$	\\
	&	$\Delta z_2$	&	30.50	\%	&	12.09	\%	&	0.21	$\sigma$	&	0.09	$\sigma$	\\
	&	$\Delta z_3$	&	14.89	\%	&	13.46	\%	&	0.37	$\sigma$	&	0.62	$\sigma$	\\
	&	$\Delta z_4$	&	7.38	\%	&	3.40	\%	&	0.44	$\sigma$	&	0.56	$\sigma$	\\
	&	$A_1$	&	55.94	\%	&	31.02	\%	&	0.89	$\sigma$	&	1.79	$\sigma$	\\
\hline															
\end{tabular}
\caption{The improvement and shift in the determination of the source redshift distribution parameters, $\Delta z$, and the amplitude of the intrinsic alignments, $A_1$, after adding SR information to the cosmic shear dataset for each weak lensing survey in the baseline set-up.  The improvements are expressed as a percentage, where the shift is expressed in units of the parameter error determined by cosmic shear alone.  Results are shown both including and excluding the prior in $\Delta z$ adopted by each weak lensing collaboration.}
\label{table:SR_improve_shift}
\end{table*}

\begin{figure*}
    \centering
    \includegraphics[width=0.49\linewidth]{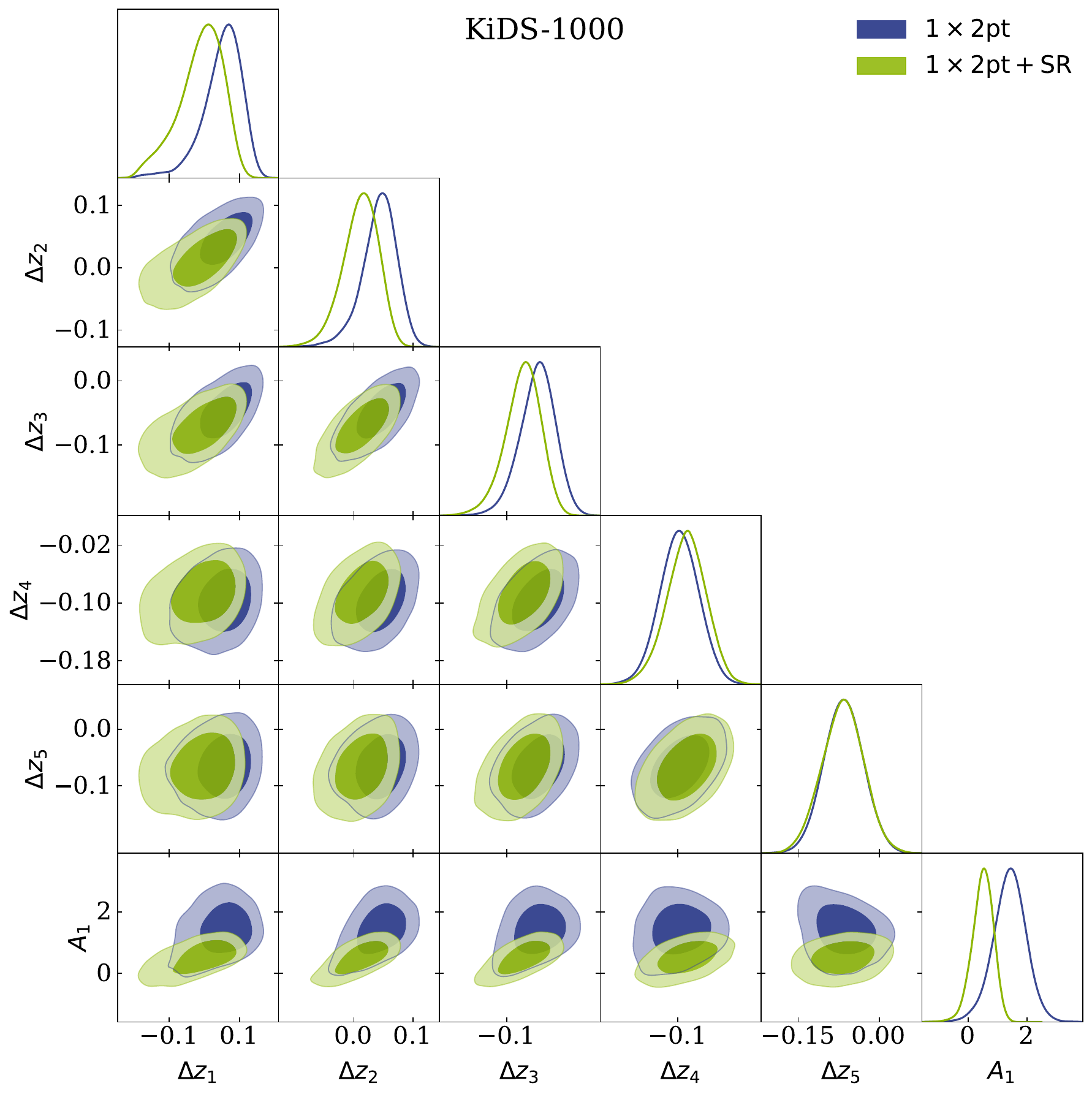}
    \includegraphics[width=0.49\linewidth]{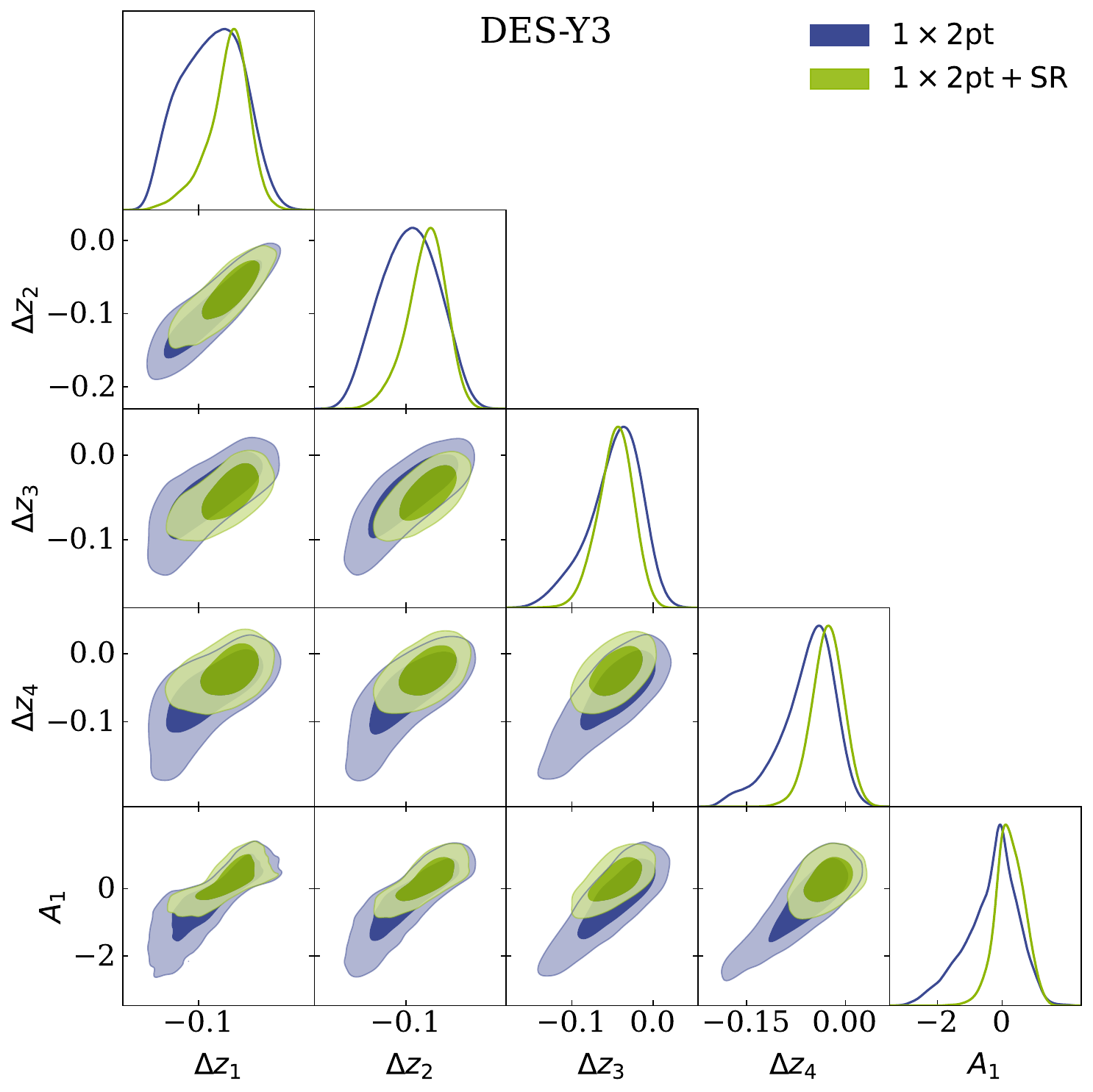}
    \vspace{1\baselineskip}
    \includegraphics[width=0.49\linewidth]{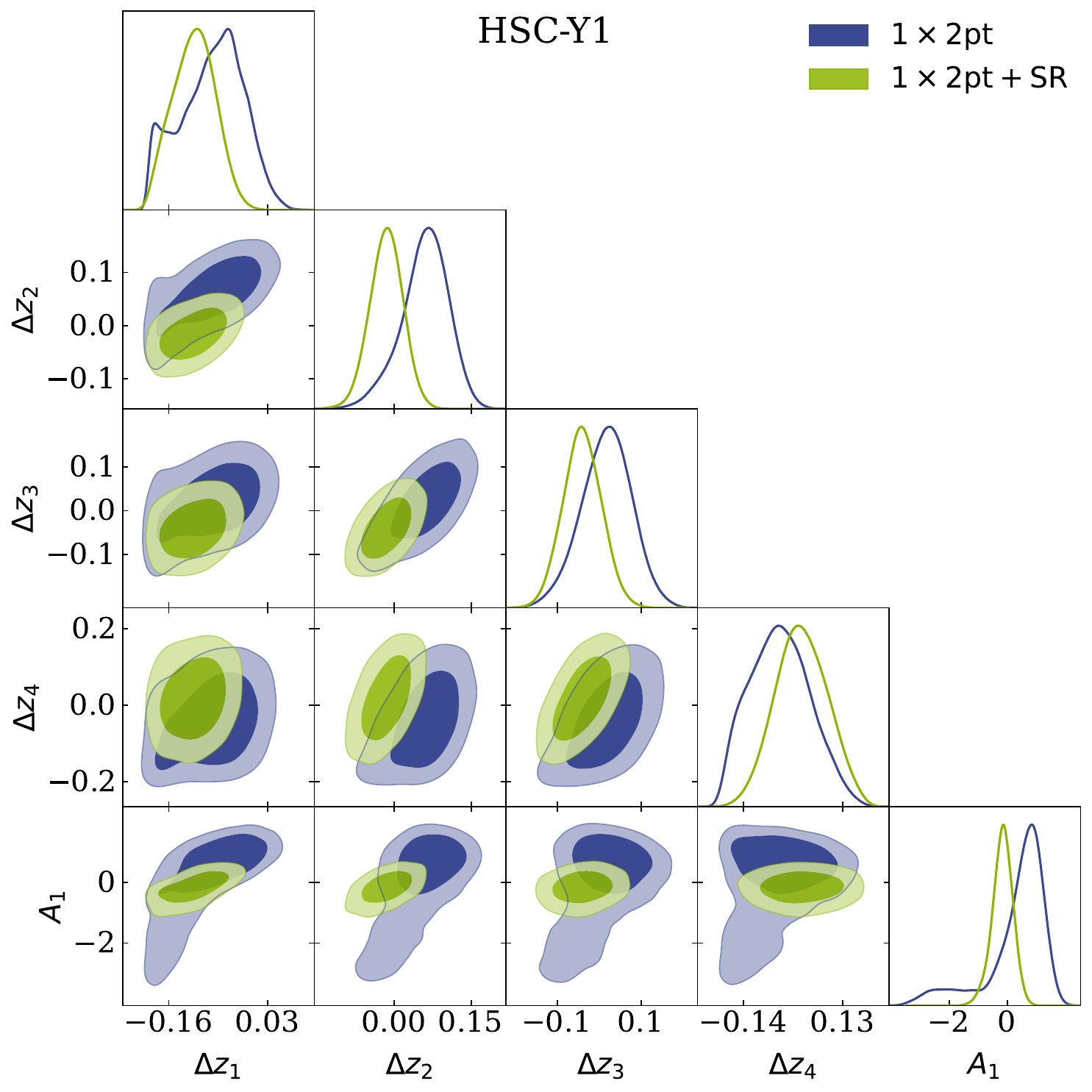}
    \includegraphics[width=0.49\linewidth]{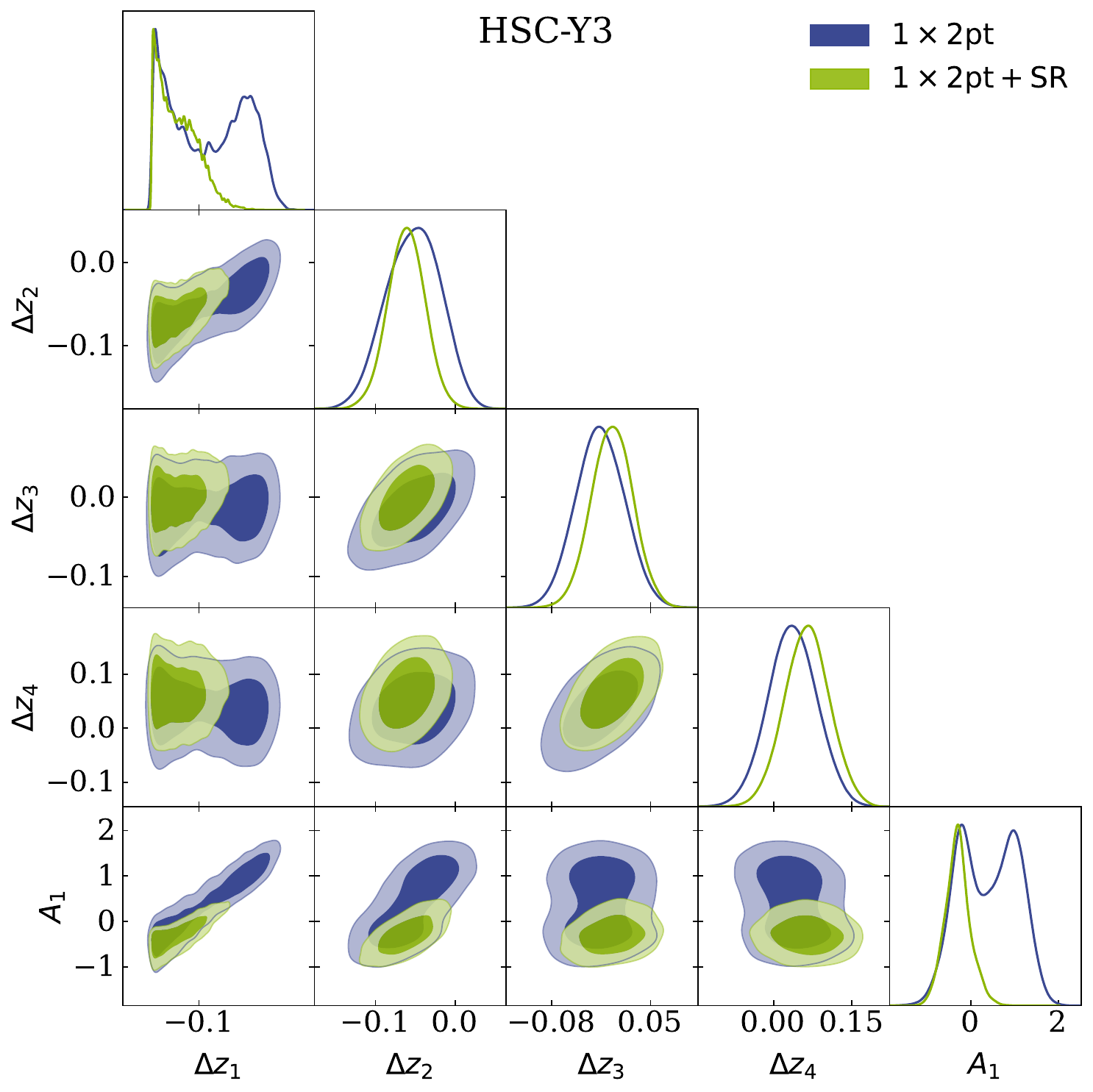}
    \caption{Joint confidence regions for fits to the intrinsic alignment amplitude $A_1$ and source redshift distribution parameters $\Delta z$ for cosmic shear (blue contours), and with the addition of shear ratio measurements (green contours). We do not include the Gaussian priors in $\Delta z$, and show results for KiDS-1000, DES-Y3, HSC-Y1 and HSC-Y3.}
    \label{corner_plot_noprior}
\end{figure*}

We now present the parameter constraints for our baseline analysis, in which we assume the DESI fiducial cosmology, the \textsc{hmcode2020} non-linear matter power spectrum model with baryon feedback, and average the shear ratio over scales $2-6 \, h^{-1}$ Mpc.  We investigate the consistency of the SR measurements with the fit to the cosmic shear data, the improvements in the determination of the astrophysical parameters from adding SR information to cosmic shear data, and the dependence of these results on the $\Delta z$ prior assumed by the weak lensing collaborations.

First, we discuss the results without the $\Delta z$ prior (all our results are displayed in Table \ref{table:SR_improve_shift}).  The DES-Y3, HSC-Y1 and HSC-Y3 analyses all show a significant improvement in the $\Delta z$ constraint after adding SR information to cosmic shear: up to 41\% for DES-Y3, 33\% for HSC-Y1 and 49\% for HSC-Y3.  Although the improvement is less significant for KiDS-1000, $\Delta z_1$ and $\Delta z_2$ nonetheless shift towards the value $\Delta z = 0$, and are no longer limited by the wide uniform prior ($-0.2 < \Delta z < 0.2$).  Adding SR information brings particularly significant improvements in the determinations of $\Delta z_1$ for HSC-Y1 and HSC-Y3, as can be seen in Figure \ref{corner_plot_noprior}.  All surveys show improvement in the determination of the IA parameters after adding SR information, due to the breaking of parameter degeneracies, as shown in Figure \ref{corner_plot_noprior}.  We can see from this figure that there are significant degeneracies between the IA amplitude, $A_1$, and the offset in the redshift distributions of $\Delta z$, causing these parameters to be weakly constrained in the absence of a $\Delta z$ prior, and the $A_1$ posterior to be wide and indeed multimodal in some cases.

We now discuss the results assuming the priors in $\Delta z$ provided by each weak lensing collaboration.  Naturally, these priors significantly reduce the improvement in the determination of the $\Delta z$ parameters from adding SR information to cosmic shear; however, we still obtain significant gains in the measurement of the IA parameters, in the range $13-34\%$.  We also note that in the case of HSC-Y1 and HSC-Y3, the SR measurements favour somewhat lower values of the amplitude of intrinsic alignments, $A_1$, corresponding to shifts in the range $1.4-1.8 \, \sigma$.  These findings are related to the assumed cosmology, as we will discuss in the following section.

Figure \ref{Fid_best_SR} displays the best-fitting models for the angle-averaged shear ratio resulting from the joint fits to the combined shear+SR dataset.  In all cases, the model provides a good description of the SR measurements, with $\chi^2$ values for this part of the data vector listed in Table \ref{tab:chisq}.  This provides validation that our modelling framework successfully describes the shear ratio measurements of all lensing datasets.

\subsubsection{Variations of the analysis choices}

\begin{table}
\centering
\begin{tabular}{c c c c c }
        \hline
        Data & $\chi^2$(all) & $\chi^2$(SR) & $N_{\textrm{data}}$(all) & $N_{\textrm{data}}$(SR)  \\
        \hline
	\textbf{KiDS-1000}											\\
Baseline	&	289.570	&	12.400	&	249 & 24	\\
$1 - 10 \, h^{-1}$ Mpc 	&	299.699	&	18.277	&	249	&  24\\
All	scales	&	311.044	&	27.518	&	249	& 24 	\\
\textsc{hmcode2020}	&	307.674	&	13.705	&	249	& 24 	\\
\textsc{halofit}	&	313.934	&	12.215	&	249	& 24 	\\
WL cosmology	&	286.845	&	11.828	&	249	& 24 	\\
											
\hline											
\textbf{DES-Y3	}			\\							
Baseline	&	309.628	&	23.672	&	245	& 18 	\\
$1 - 10 \, h^{-1}$ Mpc 	&	300.848	&	14.751	&	245	& 18	\\
All	scales	&	307.699	&	20.414	&	245	& 18 	\\
\textsc{hmcode2020}	&	311.940	&	22.921	&	245	& 18 \\
\textsc{halofit}	&	312.659	&	23.662	&	245	& 18	\\
WL cosmology	&	315.958	&	26.059	&	245	& 18 		\\
											
\hline										
\textbf{HSC-Y1	}		\\								
Baseline	&	317.766	&	24.032	&	188	& 18 	\\
$1 - 10 \, h^{-1}$ Mpc 	&	328.760	&	33.093	&	188	& 18 	\\
All	scales &	351.211	&	51.275	&	188	& 18 	\\
\textsc{hmcode2020}	&	318.280	&	25.539	&	188	& 18 	\\
\textsc{halofit}	&	317.142	&	23.070	&	188	& 18 	\\
WL cosmology	&	320.218	&	23.997	&	188	& 18 	\\
											
\hline											
\textbf{HSC-Y3}											\\
Baseline	&	249.23	&	18.238	&	158	& 18	\\
$1 - 10 \, h^{-1}$ Mpc 	&	254.05	&	14.343	&	158 & 18 		\\
All	scales	&	247.95	&	11.430	&	158	& 18 	\\
\textsc{hmcode2020}	&	265.28	&	18.036	&	158	& 18 		\\
\textsc{halofit}	&	271.80	&	22.062	&	158	& 18 	\\
WL cosmology	&	246.41	&	16.953	& 	158	&	18 	\\
											
\hline
\end{tabular}
\caption{The best-fitting $\chi^2$ values, and the size of the data vector, for variations from the baseline model assumptions of averaging the shear ratio over scales $2-6 \, h^{-1}$ Mpc, using \textsc{hmcode2020} with baryon feedback for the non-linear power spectrum, and the DESI fiducial cosmology.  We show results using \textsc{hmcode2020} without baryon feedback and \textsc{halofit}, averaging over $1-10 \, h^{-1}$ Mpc or all separations; and switching to the best-fitting cosmology of each weak lensing (WL) collaboration.  We show results for the full data vector (``all''), and just for the shear ratio measurements (``SR'').}
\label{tab:chisq}
\end{table}

\begin{figure*}
    \centering
    \includegraphics[width=0.49\linewidth]{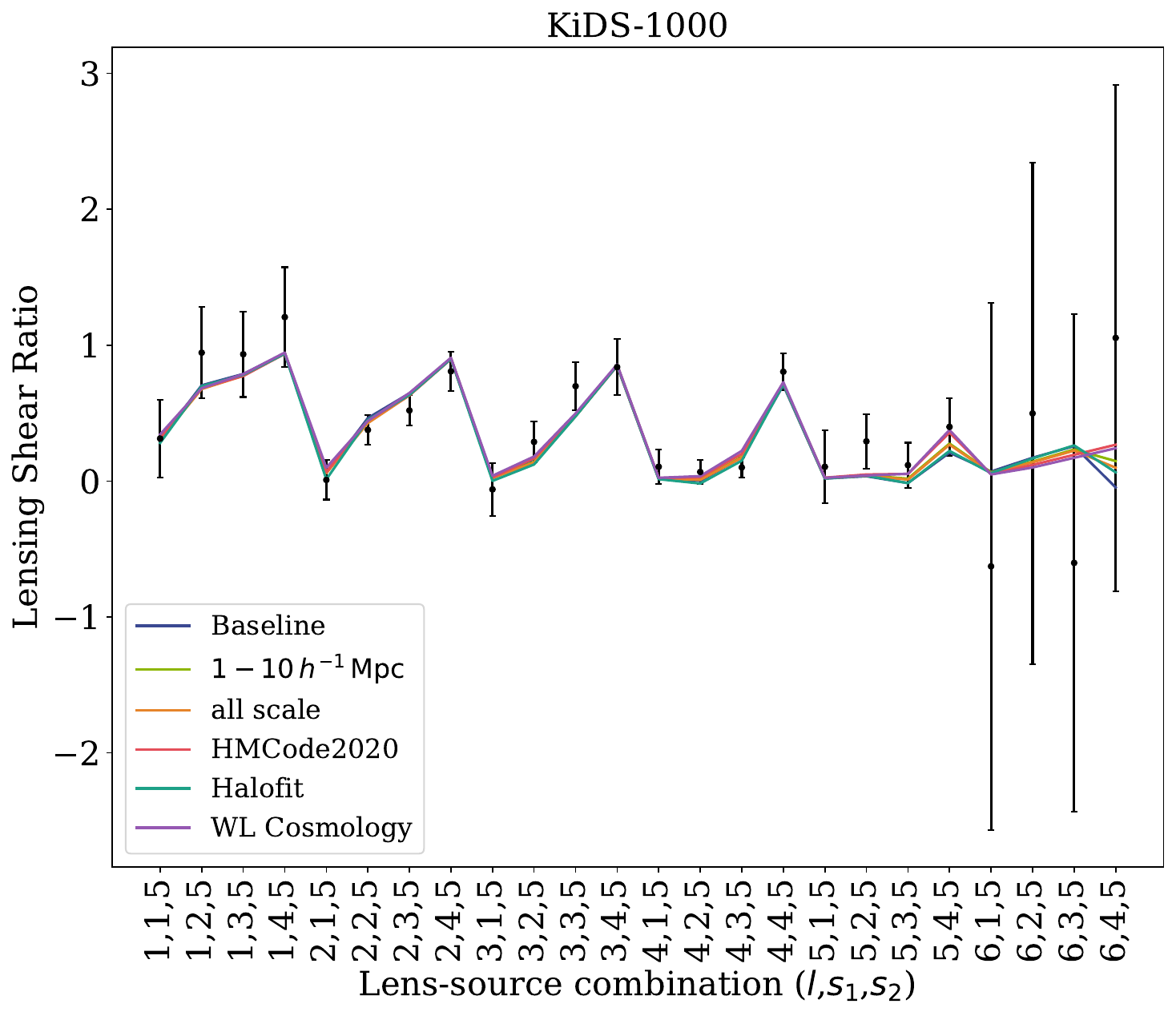}
    \includegraphics[width=0.49\linewidth]{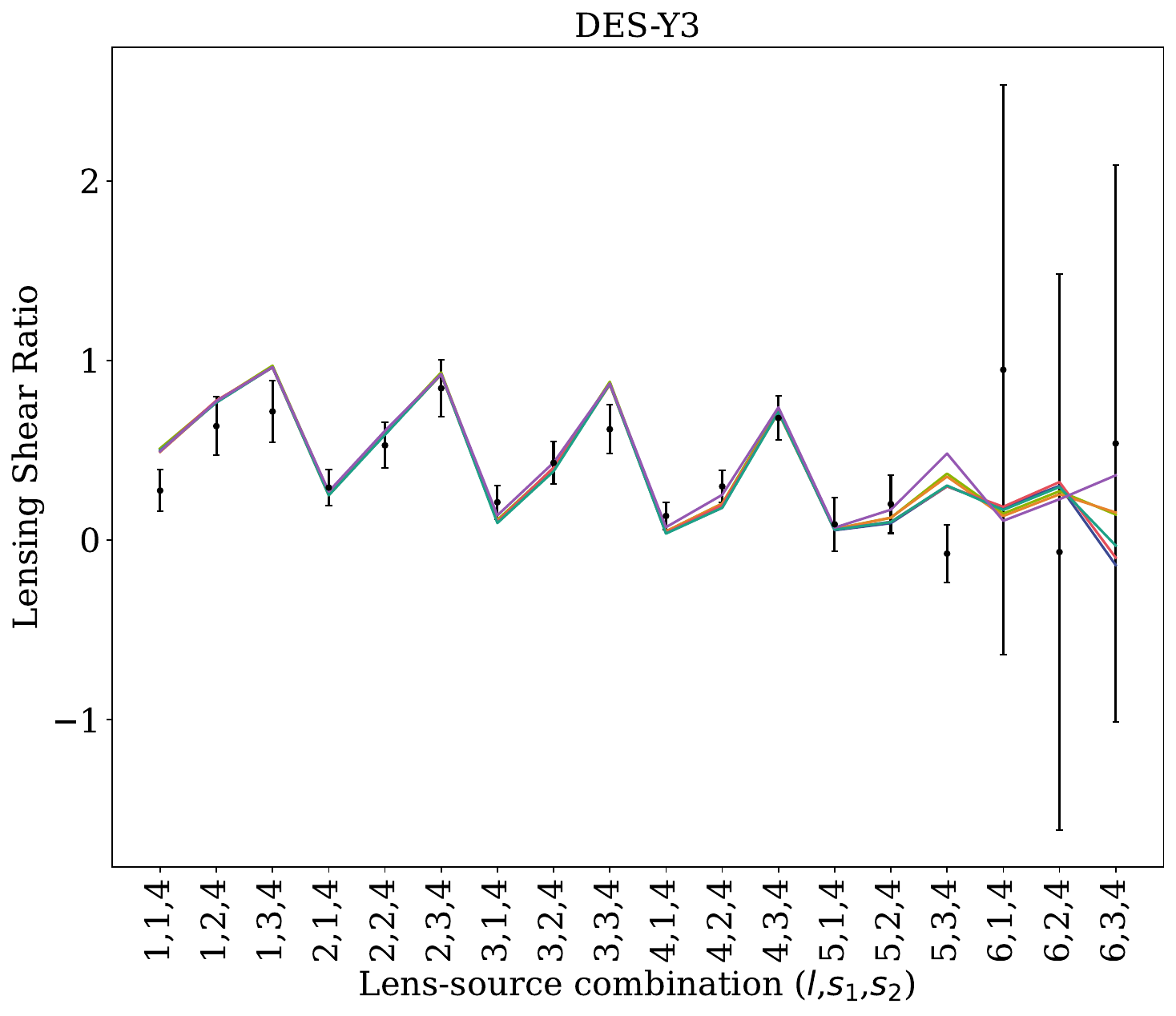}
    \vspace{1\baselineskip}
    \includegraphics[width=0.49\linewidth]{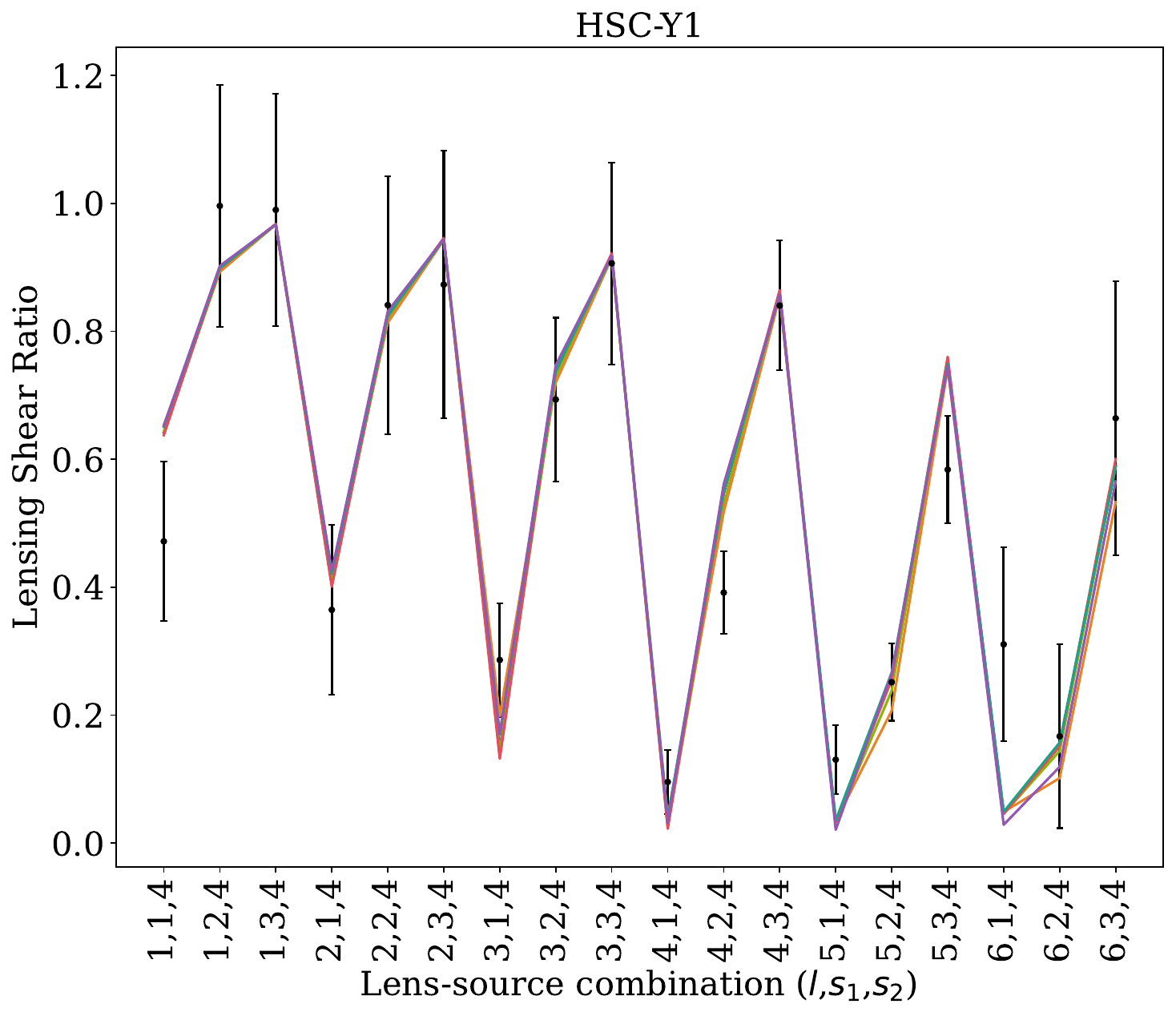}
    \includegraphics[width=0.49\linewidth]{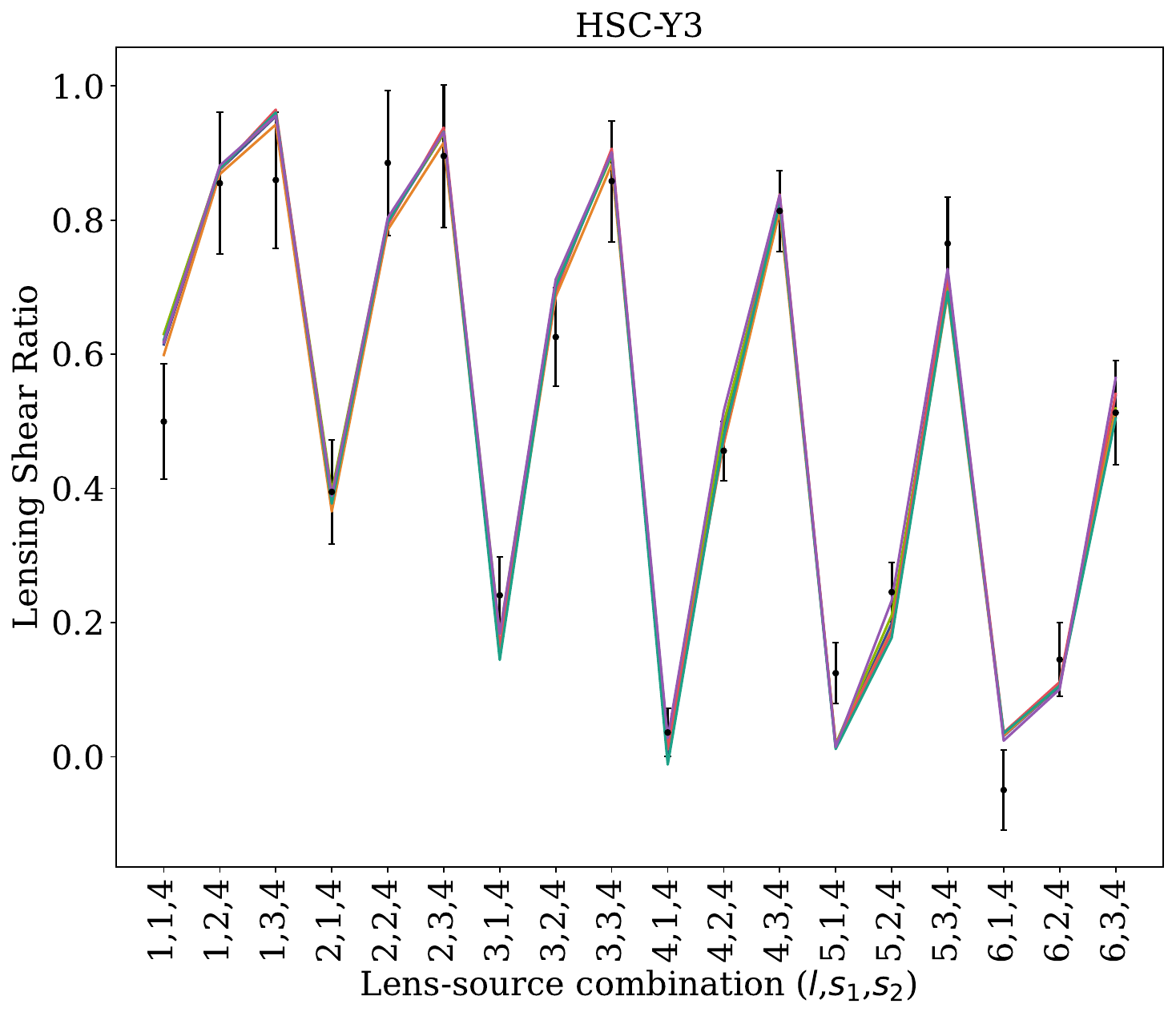}
    \caption{The measurements of the angle-averaged shear ratio from DESI-DR1 lenses and KiDS, DES-Y3, HSC-Y1 and HSC-Y3 sources, displayed as in Figure \ref{Fid_best_SR}.  In this case, we overplot the best-fitting models corresponding to a range of analysis choices, where the small variation in these models indicates that our results are robust to changing analysis frameworks.}
    \label{all_model_SR}
\end{figure*}

\begin{figure*}
  \includegraphics[width=\linewidth]{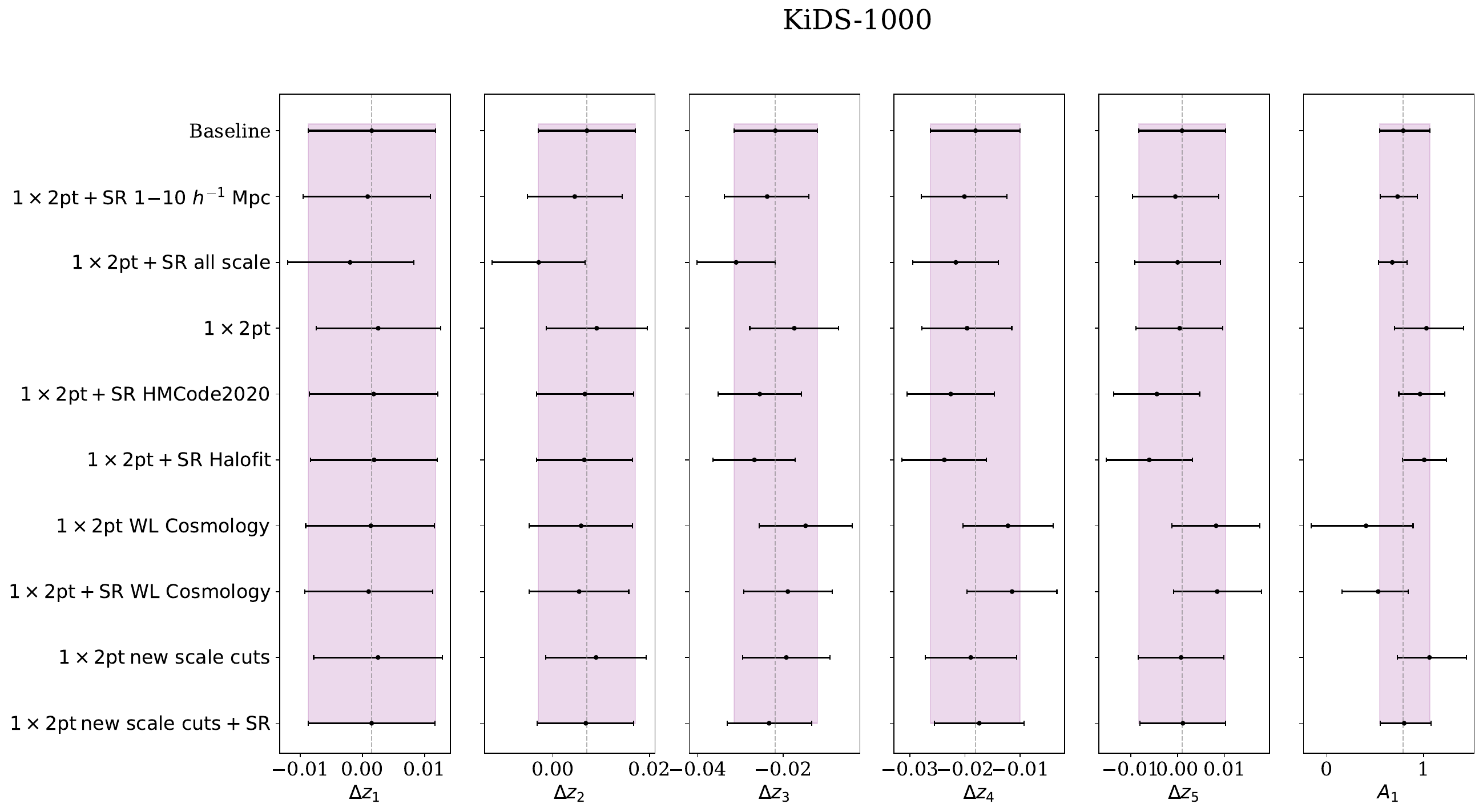}
  \caption{Comparison of the marginalised parameter constraints for the intrinsic alignment amplitude $A_1$ and source redshift distribution parameters $\Delta z$ for different cases analysed in our study, where we show results for the KiDS-1000 survey in this figure. The baseline analysis corresponds to cosmic shear combined with SR with scale $2-6 h^{-1} \rm Mpc$, and uses \textsc{hmcode2020} with baryons for the matter power spectrum model and DESI fiducial cosmology. The different rows correspond to different combinations of data and modelling choices, and the shaded region indicates our baseline fits.}
  \label{fig:bar_SR_kids}
\vspace{0.5cm}
\end{figure*}

We now discuss the results of varying the baseline model choices in the analysis.  First, we consider changing the separation range over which the SR measurements are averaged from our baseline choice of $2-6 \, h^{-1}$ Mpc to either $1-10 \, h^{-1}$ Mpc or utilising all separations.  The variation in the best-fitting models corresponding to these cases is illustrated in Figure \ref{all_model_SR}, and the changes in the minimum $\chi^2$ statistic can be seen in Table \ref{tab:chisq}.  We find that changing the separation range does not have a significant impact on the results.

We also tested varying the non-linear matter power spectrum model from our baseline choice of using \textsc{hmcode2020} with baryon feedback, to either using \textsc{hmcode2020} without baryon feedback, or \textsc{halofit}; and changing our fixed cosmological model from the DESI fiducial model to the best fit reported by each weak lensing survey.  Again, Figure \ref{all_model_SR} shows that we cannot distinguish the best-fitting models.

We note that when we change cosmological model, the best-fitting values of $\Delta z$ and the IA parameters do vary somewhat, because the distance-redshift relation and IA modelling are dependent on cosmological parameters such as $\Omega_\mathrm{m}$.  We show the comparisons of these results in Figure \ref{fig:bar_SR_kids} for KiDS-1000, where noticeable shifts are observed in the $\Delta z_4$, $\Delta z_5$ and IA parameters when changing our base cosmology.  Hence, changing the cosmological model has the result of reducing the IA parameter shift we observe when adding SR information to the cosmic shear dataset to $0.35 \, \sigma$ and $0.88 \, \sigma$ for HSC-Y1 and HSC-Y3, respectively, within the statistical error margin.  Finally, we note that switching to the revised scale cut for KiDS-1000 we derived in Section \ref{sec:Model_valid} creates no significant changes in results.  This is also shown in Figure \ref{fig:bar_SR_kids}.  For convenience, the corresponding fits for the DES-Y3, HSC-Y1 and HSC-Y3 datasets are displayed in the Appendix in Figures \ref{fig:bar_SR_desy3}, \ref{fig:bar_SR_hscy1} and \ref{fig:bar_SR_hscy3}.

\section{Conclusion}
\label{sec:conc}

In this study we have performed validation tests of the joint $3 \times 2$-pt cosmological analysis of the DESI-DR1 spectroscopic dataset and the KiDS-1000, DES-Y3, HSC-Y1 and HSC-Y3 weak lensing surveys.  In particular, we focussed on two types of validation: the selection of appropriate scale cuts to ensure that cosmological parameter fits are robust against astrophysical modelling choices, and the performance of shear ratio tests to check the internal consistency of the data in the presence of potential weak lensing systematic effects.

First, we designed scale cuts for galaxy-galaxy lensing and galaxy clustering measurements, to ensure that any cosmological parameter shifts associated with modelling of the non-linear matter power spectrum, non-linear galaxy bias, and baryon feedback are sub-dominant to statistical errors.  We followed the method of \cite{2021arXiv210513548K} which compares parameter fits to a baseline noise-free model and a ``contaminated'' model including the aforementioned systematics.  We adopted threshold criteria that the parameter shifts in $\Omega_\mathrm{m}$ and $S_8$ should be less than $30\%$ of the statistical errors in these parameters.  We used the fiducial cosmic shear scale cuts of each lensing survey, except for KiDS-1000 where we propose a minor change in the inclusion of $\xi_-$ data points for some tomographic bins \citep[using the $\chi^2$ method of][]{2021arXiv210513548K}.  We ensured that our proposed scale cuts satisfied our required parameter shift threshold when models are fit to cosmic shear only, to $2 \times2$-pt correlations (galaxy-galaxy lensing and galaxy clustering) and to the full set of $3 \times2$-pt correlations.  After testing different scale cut options, we found that $(6, 8) \, h^{-1}\mathrm{Mpc}$, for galaxy-galaxy lensing and galaxy clustering respectively, satisfied our criteria.

Second, we performed a shear ratio test using DESI-DR1 and the three weak lensing surveys.  We measured shear ratios using Bayesian methods, using the highest-redshift source tomographic bin in the denominator for all cases because it has the highest signal-to-noise ratio measurement across the full set of lens bins.  We followed the methods of \cite*{2022PhRvD.105h3529S} and \cite{Emas_SR}, fitting a full $\gamma_\mathrm{t}$ model to the measurements.  We performed joint fits to the shear ratio and cosmic shear correlations, for the source redshift distribution offsets $\Delta z$, and intrinsic alignment parameters, for fixed cosmology, and checked that our results were consistent when we varied the scales fitted in the shear ratio test, the power spectrum model and the fiducial cosmological model.  We found that all shear ratios could be successfully fit with consistent models to the cosmic shear, as tested by the $\chi^2$ statistic between the shear ratio model and data, validating our correlation modelling.

We also investigated the extent to which the shear ratio measurements provided new information to the cosmic shear.  When we did not include the $\Delta z$ prior adopted by weak lensing surveys, we found that the inclusion of shear ratio significantly improves the measurements of these astrophysical parameters, providing complementary information to the cosmic shear.  When including this prior, we found that the shear ratio still yielded significant improvements in the determination of IA parameters.  For our fixed cosmology fits, we found that the shear ratio favoured lower values of the IA amplitude in the case of HSC (by 1-2-$\sigma$), but we found that these differences could be attributed to a degeneracy with the assumed value of $\Omega_\mathrm{m}$, so would be absorbed by varying-cosmology fits.

Our work serves as a critical precursor to the full $3 \times 2$-pt cosmological analysis of these datasets, which will be reported by \cite{2025arXiv251215960P}.  These datasets will expand significantly in the future due to observations from the \textit{Euclid} Satellite \citep{2011arXiv1110.3193L, 2025A&A...697A...1E}, the Vera Rubin Observatory \citep{2019ApJ...873..111I}, the Roman Space Telescope \citep{2015arXiv150303757S, 2021MNRAS.507.1746E}, delivering significant improvements in statistical constraining power.  Further modelling and validation improvements will hence be needed, especially to study effects on non-linear scales such as baryon feedback, non-linear galaxy bias and the non-linear matter power spectrum.  With more data, we can also better understand astrophysical parameters, such as the source redshift distribution and intrinsic alignments. The shear ratio test can continue to play a significant role in testing these models, enabling ever-more precise and accurate tests of the cosmological model.

\section*{Acknowledgements}

NE would like to acknowledge the financial support received through a Swinburne University Postgraduate Research Award. AP acknowledges support from the European Union's Marie Skłodowska-Curie grant agreement 101068581, and from the \textit{César Nombela} Research Talent Attraction grant from the Community of Madrid (Ref. 2023-T1/TEC-29011). CB acknowledges financial support received through Australian Research Council Discovery Project DP220101609.

This material is based upon work supported by the U.S. Department of Energy (DOE), Office of Science, Office of High-Energy Physics, under Contract No. DE–AC02–05CH11231, and by the National Energy Research Scientific Computing Center, a DOE Office of Science User Facility under the same contract. Additional support for DESI was provided by the U.S. National Science Foundation (NSF), Division of Astronomical Sciences under Contract No. AST-0950945 to the NSF’s National Optical-Infrared Astronomy Research Laboratory; the Science and Technology Facilities Council of the United Kingdom; the Gordon and Betty Moore Foundation; the Heising-Simons Foundation; the French Alternative Energies and Atomic Energy Commission (CEA); the National Council of Humanities, Science and Technology of Mexico (CONAHCYT); the Ministry of Science, Innovation and Universities of Spain (MICIU/AEI/10.13039/501100011033), and by the DESI Member Institutions: \url{https://www.desi.lbl.gov/collaborating-institutions}. Any opinions, findings, and conclusions or recommendations expressed in this material are those of the author(s) and do not necessarily reflect the views of the U. S. National Science Foundation, the U. S. Department of Energy, or any of the listed funding agencies.

The authors are honored to be permitted to conduct scientific research on Iolkam Du’ag (Kitt Peak), a mountain with particular significance to the Tohono O’odham Nation.

\section*{Data availability}

Data points for all the figures are available at \url{https://doi.org/10.5281/zenodo.17115282}.

\bibliographystyle{mnras}
\bibliography{template}

\section*{Affiliations}
\scriptsize
\noindent
$^{1}$ Centre for Astrophysics \& Supercomputing, Swinburne University of Technology, P.O. Box 218, Hawthorn, VIC 3122, Australia\\
$^{2}$ CIEMAT, Avenida Complutense 40, E-28040 Madrid, Spain\\
$^{3}$ Institute for Astronomy, University of Edinburgh, Royal Observatory, Blackford Hill, Edinburgh EH9 3HJ, UK\\
$^{4}$ Ruhr University Bochum, Faculty of Physics and Astronomy, Astronomical Institute (AIRUB), German Centre for Cosmological Lensing, 44780 Bochum, Germany\\
$^{5}$ The Ohio State University, Columbus, 43210 OH, USA\\
$^{6}$ Physics Department, Brookhaven National Laboratory, Upton, NY 11973, USA\\
$^{7}$ Lawrence Berkeley National Laboratory, 1 Cyclotron Road, Berkeley, CA 94720, USA\\
$^{8}$ Department of Physics, Boston University, 590 Commonwealth Avenue, Boston, MA 02215 USA\\
$^{9}$ Dipartimento di Fisica ``Aldo Pontremoli'', Universit\`a degli Studi di Milano, Via Celoria 16, I-20133 Milano, Italy\\
$^{10}$ INAF-Osservatorio Astronomico di Brera, Via Brera 28, 20122 Milano, Italy\\
$^{11}$ Department of Physics \& Astronomy, University College London, Gower Street, London, WC1E 6BT, UK\\
$^{12}$ Institut d'Estudis Espacials de Catalunya (IEEC), c/ Esteve Terradas 1, Edifici RDIT, Campus PMT-UPC, 08860 Castelldefels, Spain\\
$^{13}$ Institute of Space Sciences, ICE-CSIC, Campus UAB, Carrer de Can Magrans s/n, 08913 Bellaterra, Barcelona, Spain\\
$^{14}$ Instituto de F\'{\i}sica, Universidad Nacional Aut\'{o}noma de M\'{e}xico,  Circuito de la Investigaci\'{o}n Cient\'{\i}fica, Ciudad Universitaria, Cd. de M\'{e}xico  C.~P.~04510,  M\'{e}xico\\
$^{15}$ NSF NOIRLab, 950 N. Cherry Ave., Tucson, AZ 85719, USA\\
$^{16}$ Department of Astronomy \& Astrophysics, University of Toronto, Toronto, ON M5S 3H4, Canada\\
$^{17}$ Department of Physics \& Astronomy and Pittsburgh Particle Physics, Astrophysics, and Cosmology Center (PITT PACC), University of Pittsburgh, 3941 O'Hara Street, Pittsburgh, PA 15260, USA\\
$^{18}$ University of California, Berkeley, 110 Sproul Hall \#5800 Berkeley, CA 94720, USA\\
$^{19}$ Departamento de F\'isica, Universidad de los Andes, Cra. 1 No. 18A-10, Edificio Ip, CP 111711, Bogot\'a, Colombia\\
$^{20}$ Observatorio Astron\'omico, Universidad de los Andes, Cra. 1 No. 18A-10, Edificio H, CP 111711 Bogot\'a, Colombia\\
$^{21}$ Center for Astrophysics $|$ Harvard \& Smithsonian, 60 Garden Street, Cambridge, MA 02138, USA\\
$^{22}$ NASA Einstein Fellow\\
$^{23}$ Institute of Cosmology and Gravitation, University of Portsmouth, Dennis Sciama Building, Portsmouth, PO1 3FX, UK\\
$^{24}$ University of Virginia, Department of Astronomy, Charlottesville, VA 22904, USA\\
$^{25}$ Fermi National Accelerator Laboratory, PO Box 500, Batavia, IL 60510, USA\\
$^{26}$ Department of Astronomy and Astrophysics, UCO/Lick Observatory, University of California, 1156 High Street, Santa Cruz, CA 95064, USA\\
$^{27}$ Center for Cosmology and AstroParticle Physics, The Ohio State University, 191 West Woodruff Avenue, Columbus, OH 43210, USA\\
$^{28}$ Department of Physics, The Ohio State University, 191 West Woodruff Avenue, Columbus, OH 43210, USA\\
$^{29}$ Department of Physics, University of Michigan, 450 Church Street, Ann Arbor, MI 48109, USA\\
$^{30}$ University of Michigan, 500 S. State Street, Ann Arbor, MI 48109, USA\\
$^{31}$ Department of Physics, The University of Texas at Dallas, 800 W. Campbell Rd., Richardson, TX 75080, USA\\
$^{32}$ Aix Marseille Univ, CNRS, CNES, LAM, Marseille, France\\
$^{33}$ Department of Physics, Southern Methodist University, 3215 Daniel Avenue, Dallas, TX 75275, USA\\
$^{34}$ Department of Physics and Astronomy, University of California, Irvine, 92697, USA\\
$^{35}$ Department of Physics and Astronomy, University of Waterloo, 200 University Ave W, Waterloo, ON N2L 3G1, Canada\\
$^{36}$ Perimeter Institute for Theoretical Physics, 31 Caroline St. North, Waterloo, ON N2L 2Y5, Canada\\
$^{37}$ Waterloo Centre for Astrophysics, University of Waterloo, 200 University Ave W, Waterloo, ON N2L 3G1, Canada\\
$^{38}$ Department of Physics, American University, 4400 Massachusetts Avenue NW, Washington, DC 20016, USA\\
$^{39}$ Sorbonne Universit\'{e}, CNRS/IN2P3, Laboratoire de Physique Nucl\'{e}aire et de Hautes Energies (LPNHE), FR-75005 Paris, France\\
$^{40}$ Department of Astronomy and Astrophysics, University of California, Santa Cruz, 1156 High Street, Santa Cruz, CA 95065, USA\\
$^{41}$ Departament de F\'{i}sica, Serra H\'{u}nter, Universitat Aut\`{o}noma de Barcelona, 08193 Bellaterra (Barcelona), Spain\\
$^{42}$ Institut de F\'{i}sica d’Altes Energies (IFAE), The Barcelona Institute of Science and Technology, Edifici Cn, Campus UAB, 08193, Bellaterra (Barcelona), Spain\\
$^{43}$ Instituci\'{o} Catalana de Recerca i Estudis Avan\c{c}ats, Passeig de Llu\'{\i}s Companys, 23, 08010 Barcelona, Spain\\
$^{44}$ Instituto de Astrof\'{i}sica de Andaluc\'{i}a (CSIC), Glorieta de la Astronom\'{i}a, s/n, E-18008 Granada, Spain\\
$^{45}$ Department of Physics and Astronomy, Sejong University, 209 Neungdong-ro, Gwangjin-gu, Seoul 05006, Republic of Korea\\
$^{46}$ Queensland University of Technology,  School of Chemistry \& Physics, George St, Brisbane 4001, Australia\\
$^{47}$ Max Planck Institute for Extraterrestrial Physics, Gie\ss enbachstra\ss e 1, 85748 Garching, Germany\\
$^{48}$ Department of Physics \& Astronomy, Ohio University, 139 University Terrace, Athens, OH 45701, USA\\
$^{49}$ Department of Astronomy, Tsinghua University, 30 Shuangqing Road, Haidian District, Beijing, China, 100190\\
$^{50}$ Kavli Institute for Particle Astrophysics and Cosmology, Stanford University, Menlo Park, CA 94305, USA\\
$^{51}$ Physics Department, Stanford University, Stanford, CA 93405, USA\\
$^{52}$ SLAC National Accelerator Laboratory, 2575 Sand Hill Road, Menlo Park, CA 94025, USA\\
\normalsize

\appendix

\section{Parameter fits for DES-Y3, HSC-Y1 and HSC-Y3}

\begin{figure*}
  \includegraphics[width=\linewidth]{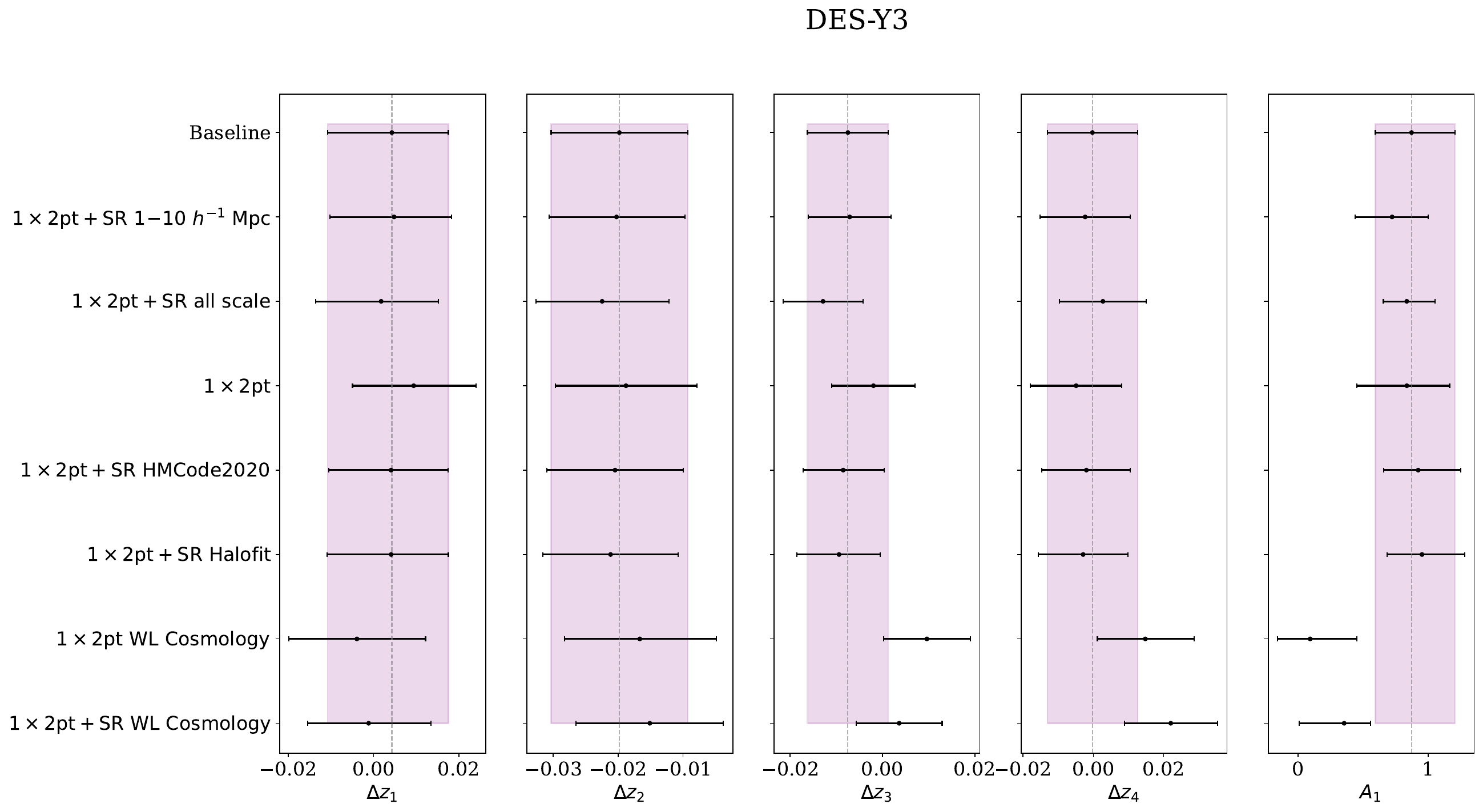}
  \caption{Comparison of the marginalised parameter constraints for the IA amplitude $A_1$ and source redshift distribution parameters $\Delta z$ for different cases analysed in our study, where we show results for DES-Y3 in this figure. Results are displayed as in Figure \ref{fig:bar_SR_kids}.}
  \label{fig:bar_SR_desy3}
\vspace{0.5cm}
\end{figure*}

\begin{figure*}
  \includegraphics[width=\linewidth]{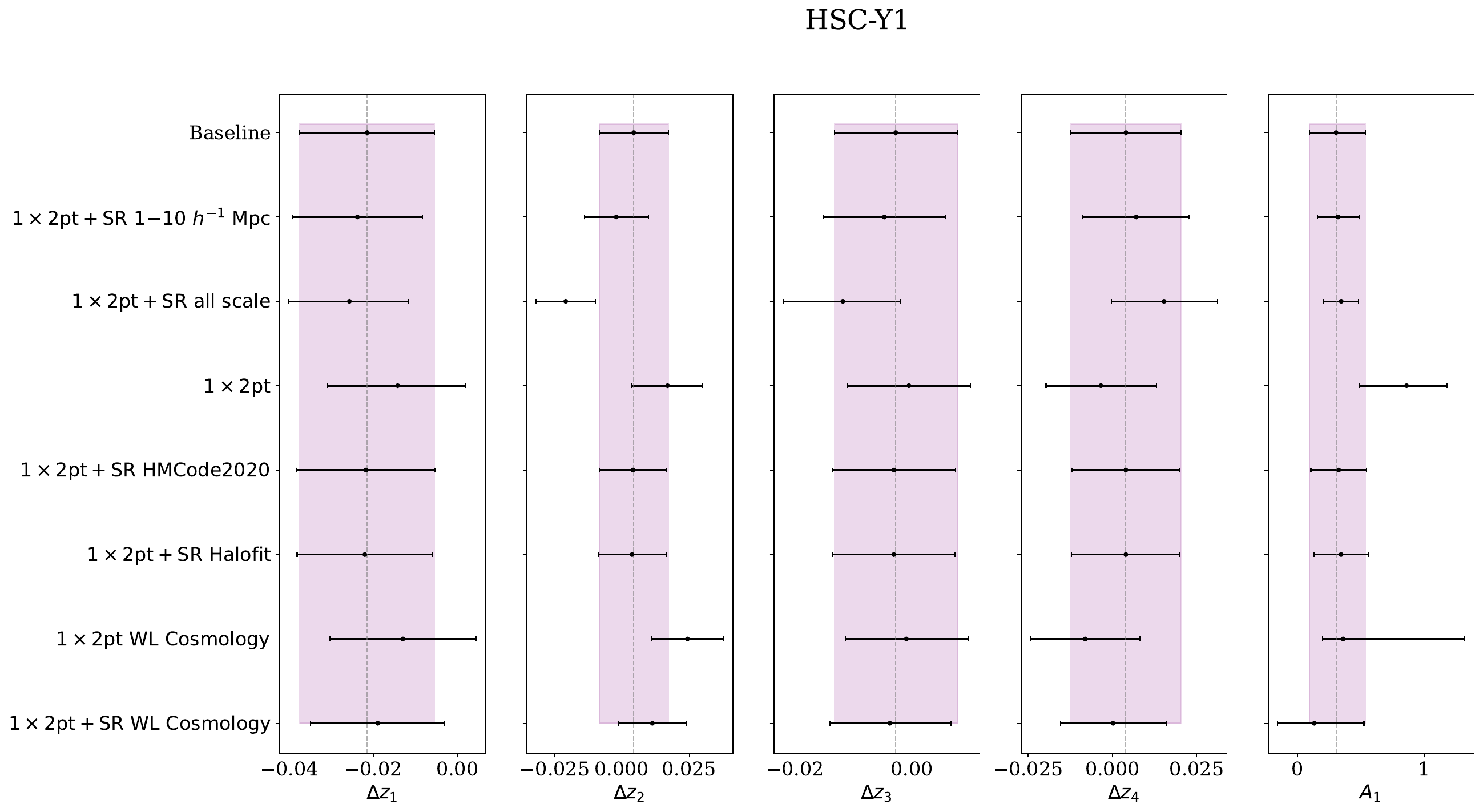}
  \caption{Comparison of the marginalised parameter constraints for the IA amplitude $A_1$ and source redshift distribution parameters $\Delta z$ for different cases analysed in our study, where we show results for HSC-Y1 in this figure. Results are displayed as in Figure \ref{fig:bar_SR_kids}.}
  \label{fig:bar_SR_hscy1}
\vspace{0.5cm}
\end{figure*}

\begin{figure*}
  \includegraphics[width=\linewidth]{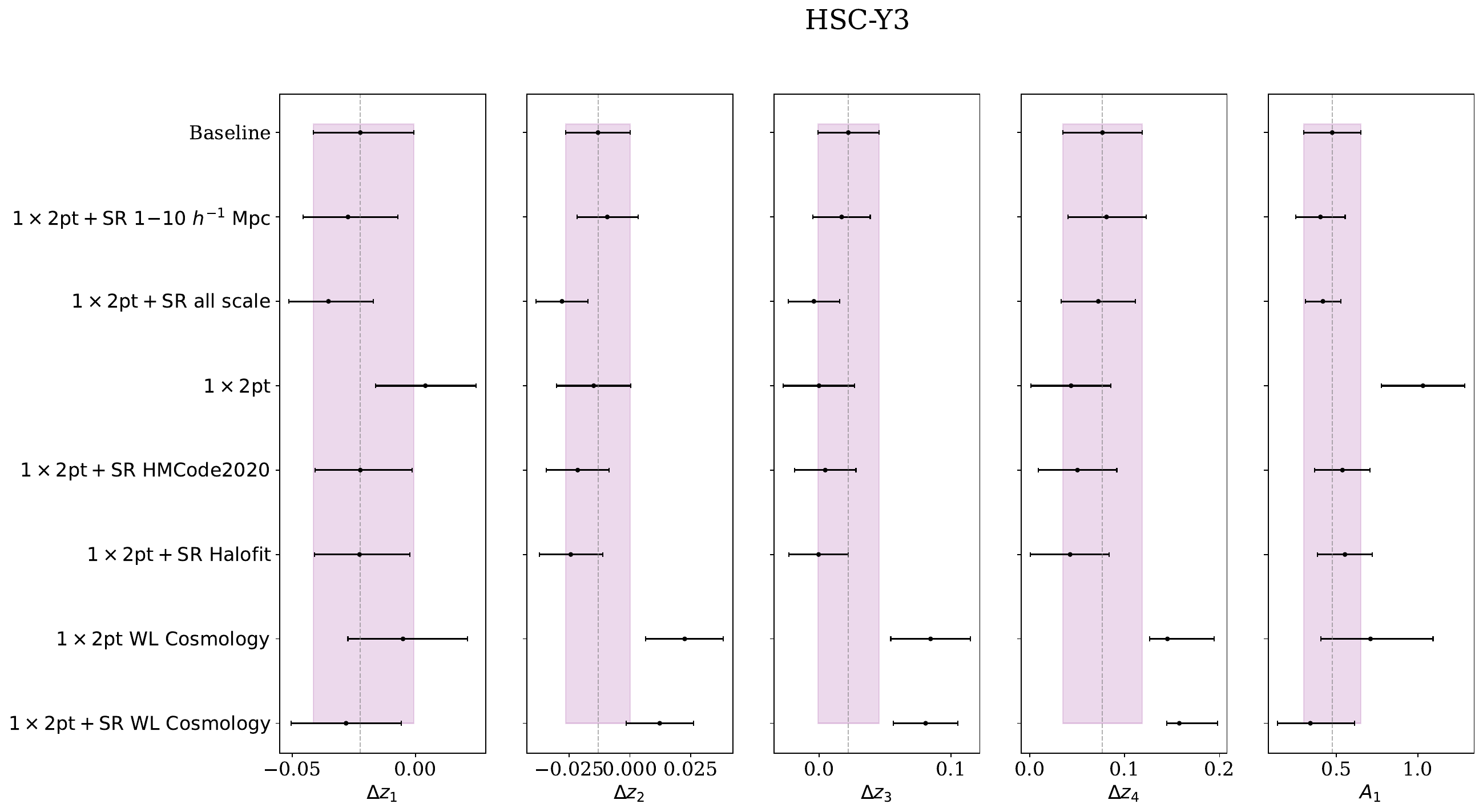}
  \caption{Comparison of the marginalised parameter constraints for the IA amplitude $A_1$ and source redshift distribution parameters $\Delta z$ for different cases analysed in our study, where we show results for HSC-Y3 in this figure. Results are displayed as in Figure \ref{fig:bar_SR_kids}.}
  \label{fig:bar_SR_hscy3}
\vspace{0.5cm}
\end{figure*}

In this Appendix we show the marginalised parameter constraints for the intrinsic alignment amplitude $A_1$ for the NLA model, and the redshift offset parameters $\Delta z$, for DES-Y3 in Figure \ref{fig:bar_SR_desy3}, HSC-Y1 in Figure \ref{fig:bar_SR_hscy1} and HSC-Y3 in Figure \ref{fig:bar_SR_hscy3}.  These complement the results already shown for KiDS-1000 in Figure \ref{fig:bar_SR_kids}.

\end{document}